\documentclass[a4paper, 10pt, final, twoside, conference]{IEEEtran} 
\usepackage{etex}
\usepackage{amsmath}
\usepackage{amssymb}
\usepackage{amsthm}
\usepackage{amsfonts}
\usepackage{mathtools} 
\usepackage{mathabx}
\usepackage{bbold}
\usepackage{thmtools}
\usepackage{tikz-cd} % for commutative diagrams
\usepackage{bussproofs} % for proof trees
\usepackage{stmaryrd} % for semantics brackets
\usepackage{microtype}
\usepackage{mdframed} % for putting boxes round things
\usepackage{hyperref}
\usepackage{wrapfig}
\usepackage{xspace}
\usepackage{xfrac}

\newenvironment{manyfigcap}
{
\begin{minipage}{\textwidth}
\centering
\vspace{8pt}
}
{
\end{minipage}
}

\newcommand{\noemph}[1]{#1}
\newcommand{\cf}{\emph{c.f.}}
\newcommand{\Def}[1]{\emph{#1}}
\newcommand{\noDef}[1]{#1}

\newcommand{\eg}{\emph{e.g.}}
\newcommand{\ie}{\emph{i.e.}}

\newcommand{\st}{\mid} % such that
\newcommand{\rulename}[1]{{\tt #1}}

\newcommand{\To}{\ensuremath{\Rightarrow}} % alias for \Rightarrow
\newcommand{\XRA}[1]{\ensuremath{\xRightarrow{#1}}}
\newcommand{\id}{\ensuremath{\mathrm{id}}} % for roman id
\newcommand{\Id}{\mathrm{Id}} % for Id 
\newcommand{\eval}{\mathrm{eval}} % for roman eval

\newcommand{\altToHead}{\raisebox{.28mm}{\scriptsize$\rhd$}}%{\triangleright}
\DeclareMathOperator{\altTo}{\mathtt{=\!}\altToHead}
\DeclareMathOperator{\scriptsizealtTo}{\mathtt{=\!}\scriptsizealtToHead}
\newcommand{\scriptsizealtToHead}{\raisebox{.21mm}{\tiny$\rhd$}}%{\triangleright}

\newcommand{\expobj}[2]{#1 \altTo #2} % exponential
\newcommand{\scriptsizeexpobj}[2]{#1 \scriptsizealtTo #2} % exponential
\newcommand{\iso}{\cong}
\newcommand{\seq}[1]{\ensuremath{[#1]}} % for [ .. ]
\newcommand{\tup}[1]{\ensuremath{\langle #1 \rangle}} % for < .. >
\newcommand{\ind}[1]{{#1}_{\bullet}} % for indexed lists / sequences

\newcommand{\sem}[1]{{\ensuremath{\llbracket #1 \rrbracket}}} % for things in semantics brackets
\newcommand{\semext}{^{\#}} % for things in semantics brackets
\newcommand{\adjUp}{{\mathbin{\rotatebox[origin=c]{270}{$\dashv$}}}}
\newcommand{\Nat}{\mathbb{N}}
\newcommand{\Hom}{\mathrm{Hom}}

\DeclareMathOperator*{\dotop}{.}
\renewcommand{\dot}{{\bf \dotop}}

\newcommand{\baseCat}{\mathcal{B}}
\newcommand{\altCat}{\mathcal{C}}

\newcommand{\catOne}{\ensuremath{\mathbb{1}}} % for the 1 category
\newcommand{\Cat}{\mathbf{Cat}}

\newcommand{\CartClosedBicat}{\mathbf{CCBicat}}
\newcommand{\termCatSymbol}{\mathcal{T}_{\mathrm{ps}}}
\newcommand{\slimCatSymbol}{\mathcal{S}_{\mathrm{ps}}}
\newcommand{\termCat}{\termCatSymbol^{\mathrm{b}}}
\newcommand{\termCatSlim}{\slimCatSymbol^{\mathrm{b}}}
\newcommand{\termCatTimes}{\termCatSymbol^{\mathsf{x}}}
\newcommand{\termCatTimesSlim}{\slimCatSymbol^{\mathsf{x}}}
\newcommand{\termCatCCC}{\termCatSymbol^{\mathsf{x},{\to}}}
\newcommand{\termCatCCCSlim}{\slimCatSymbol^{\mathsf{x},{\to}}}

\newcommand{\CartBicat}{\mathbf{fp}\text{-}\mathbf{Bicat}}
\newcommand{\Bicat}{\mathbf{Bicat}}

\newcommand{\TwoMultiGraph}{\mathrm{2}\text{-}\mathrm{MGrph}}
\newcommand{\BlankGraph}[1]{{{#1}}\text{-}\mathrm{Grph}}

\newcommand{\TwoGraph}{\BlankGraph{2}}

\renewcommand{\a}{\mathbf{a}}
\renewcommand{\r}{\mathbf{r}}
\renewcommand{\l}{\mathbf{l}}
\newcommand{\p}{\mathsf{p}}

\newcommand{\clone}{\mathcal{C}}
\newcommand{\graph}{\mathcal{G}}
\newcommand{\nodes}[1]{{#1}_0}

\newcommand{\clonesub}[2]{{#1}[{#2}]}

\newcommand{\lincore}[1]{{#1}_{\ell{c}}}
\newcommand{\cloneinto}[1]{{#1}_{c\ell}}

\newcommand{\prodop}{\textstyle{\prod}}

\newcommand{\evalterm}{\mathsf{eval}}
\newcommand{\genevalterm}[1]{\hcomp{\evalterm}{{#1}}}
\newcommand{\pair}[1]{\mathsf{tup}({#1})}
\newcommand{\proj}[1]{{\varrho_{{#1}}}}
\newcommand{\indproj}[2]{{\varrho^{(#1)}_{{#2}}}}
\newcommand{\subid}[1]{\iota_{{#1}}}
\newcommand{\assoc}[1]{\mathsf{assoc}_{{#1}}}

\newcommand{\bind}{.}

\newcommand{\lang}{\Lambda_{\mathrm{ps}}}
\newcommand{\langBicat}{\lang^{\mathrm{b}}}
\newcommand{\langCartClosed}{\lang^{{\mathsf{x},\to}}}

\newcommand{\langCart}{\lang^{\mathsf{x}}}

\newcommand{\allTypes}[1]{\mathrm{T}({#1})}%{\mathrm{T}^{\mathsf{x},\to}({#1})}
\newcommand{\allProdTypes}[1]{\mathrm{T}_0({#1})}%{\mathrm{T}^{\mathsf x}({#1})}

\newcommand{\exptype}[2]{\ensuremath{#1 \altTo #2}}
\newcommand{\unittype}{\mathbf{1}}
\newcommand{\rewrite}[2]{\ensuremath{#1 \To #2}}
\newcommand{\birewrite}[2]{\ensuremath{#1 \To #2}}

\newcommand{\lam}[2]{\lambda {#1} \dot {#2}}

\newcommand{\transExp}[1]{\mathsf{e}^{\dagger\!}(#1)}
\newcommand{\transTimes}[1]{\mathsf{p}^{\dagger\!}(#1)}
\newcommand{\transTimesSymb}{\mathsf{p}^{\dagger\!}}
\newcommand{\transExpSymb}{\mathsf{e}^{\dagger\!}}

\newcommand{\etaTimes}[1]{\varsigma_{#1}}
\newcommand{\epsilonTimesname}{\varpi}

\newcommand{\epsilonTimesInd}[2]{{\epsilonTimesname^{(#1)}_{{#2}}}}

\newcommand{\etaExp}[1]{\eta_{#1}}
\newcommand{\epsilonExpRewrName}{\epsilon}
\newcommand{\epsilonExpRewr}[1]{\epsilonExpRewrName_{#1}}
\newcommand{\genEpsilonExp}[1]{\beta_{#1}}

\newcommand{\epsilonExp}{\epsilon}

\newcommand{\wkn}[2]{\hcomp{{#1}}{\mathrm{inc}_{{#2}}}}

\newcommand{\horizComp}[2]{ {#1} \{ {#2} \} }
\newcommand{\hcomp}[2]{\horizComp{#1}{#2}}

\DeclareMathOperator{\bulletop}{\bullet}
\renewcommand{\vert}{\bulletop}
\newcommand{\vertsub}[1]{\vert}

\newcommand{\timessuper}{{\mathsf{x}}}
\newcommand{\expsuper}{{\scriptsizealtTo}}

\newcommand{\evBar}{s} % for the canonical exponential pres map
\newcommand{\prodPres}{k^\timessuper}

\newcommand{\expPres}{k^\expsuper}

\newcommand{\inc}{\iota}

\newcommand{\treeskip}{1em} % for seperating trees

\newenvironment{bprooftree}
  {\leavevmode\hbox\bgroup}
  {\DisplayProof\egroup}

\newcommand{\unaryRule}[3]{\begin{bprooftree}
\AxiomC{\ensuremath{#1}}
\RightLabel{\scriptsize #3}
\UnaryInfC{\ensuremath{#2}}
\end{bprooftree}\vspace{\treeskip}}

\newcommand{\binaryRule}[4]{\begin{bprooftree}
\AxiomC{\ensuremath{#1}}
\AxiomC{\ensuremath{#2}}
\RightLabel{\scriptsize #4}
\BinaryInfC{\ensuremath{#3}}
\end{bprooftree}\vspace{\treeskip}}

\newcommand{\trinaryRule}[5]{\begin{bprooftree}
\AxiomC{\ensuremath{#1}}
\AxiomC{\ensuremath{#2}}
\AxiomC{\ensuremath{#3}}
\RightLabel{\scriptsize #5}
\TrinaryInfC{\ensuremath{#4}}
\end{bprooftree}\vspace{\treeskip}}

\theoremstyle{plain}
\newtheorem{thm}{Theorem}[section]
\newtheorem{lemma}[thm]{Lemma}
\newtheorem{cor}[thm]{Corollary}

\newtheorem{propn}[thm]{Proposition}

\newtheorem{rule-axiom}[thm]{Rule}
\newtheorem{constr}[thm]{Construction}
\newtheorem{conj}[thm]{Conjecture}
\newtheorem*{nnconj}{Conjecture}

\theoremstyle{definition}
\newtheorem{defn}[thm]{Definition}

\theoremstyle{remark}
\newtheorem{remark}[thm]{Remark}
\newtheorem{exmp}[thm]{Example}
\newtheorem{notation}[thm]{Notation}

\newenvironment{myremark}{\begin{remark}}{\end{remark}}
\newenvironment{mylemma}{\begin{lemma}}{\end{lemma}}

\newenvironment{mythm}{\begin{thm}}{\end{thm}}
\newenvironment{mydefn}{\begin{defn}}{\end{defn}}
\newenvironment{myconstr}{\begin{constr}}{\end{constr}}
\newenvironment{mycor}{\begin{cor}}{\end{cor}}
\newenvironment{mynotation}{\begin{notation}}{\end{notation}}

\newenvironment{myexmp}{\begin{exmp}}{\end{exmp}}

\title{A type theory for cartesian closed bicategories}

\author{
\IEEEauthorblockN{\hfill Marcelo Fiore \hfill\hfill Philip Saville\hfill}
\IEEEauthorblockA{Department of Computer Science and Technology, 
  University of Cambridge} 
}

\begin{document} 

\maketitle 

\begin{IEEEkeywords}
typed lambda calculus, higher category theory, Curry-Howard-Lambek
correspondence, cartesian closed bicategories 
\end{IEEEkeywords}

\begin{abstract}
We construct an internal language for cartesian closed bicategories.  
Precisely, we introduce a type theory modelling the structure of a cartesian
closed bicategory and show that its syntactic model satisfies an appropriate
universal property, thereby lifting the Curry-Howard-Lambek correspondence to
the bicategorical setting.  Our approach is principled and practical.  Weak
substitution structure is constructed using a bicategorification of the notion
of abstract clone from universal algebra, and the rules for products and
exponentials are synthesised from semantic considerations.  The result is a
type theory that employs a novel combination of \mbox{2-dimensional} type
theory and explicit substitution, and directly generalises the Simply-Typed
Lambda Calculus.  This work is the first step in a programme aimed at proving
coherence for cartesian closed bicategories.  
\end{abstract}

\section{Introduction} 

\Def{2-categories} axiomatise the structures formed by classes of categories,
such as the \mbox{2-category} $\Cat$ of small categories, functors and natural
transformations.  In such settings the associativity and unit laws of
composition hold \noemph{strictly} (`on the nose'). In many situations---in
particular where composition is defined by a universal property---these laws
only hold up to coherent isomorphism: the resulting structure is that of a
\Def{bicategory}.  
Bicategories are rife in mathematics and theoretical computer science, arising
for instance in
algebra~\cite{Benabou1967,Street1995}, 
semantics of computation~\cite{CattaniFioreWinskel,CCRW2017}, 
datatype models~\cite{Abbott2003,Dagand2013}, 
categorical logic~\cite{FioreSpecies,Gambino2013}, and
categorical algebra~\cite{Fiore2015,Gambino2017,FGHW2017}.

The layers of coherence data required to witness the associativity and unit
laws makes calculating in bicategories (and 
weak \mbox{$n$-categories} 
more generally) notoriously difficult. One approach is to reduce bicategorical
structure to categorical structure by quotienting, but the loss of intensional
information this entails is often unsatisfactory. 

There are two main strategies for working with structures defined up to
isomorphism.  One strategy looks for \Def{coherence theorems} establishing
that some class of diagrams always commute. For bicategories and bicategorical
limits (bilimits~\cite{Gray1974}) there are well-known coherence
results~\cite{MacLane1985, Power1989bilimit}; however, we know of no analogous
result in the literature for closed structure.  Another strategy employs a
type theory that matches the categorical structure 
(see~\eg~\cite{elephant, Gambino2004, Gambino2013});
such a system is sometimes called the 
\Def{internal language}~\cite{Lambek1985} or
\Def{internal logic}~\cite{Makkai1977}. 

In this paper we carry out the \noemph{internal-language} strategy for
cartesian closed bicategories, and thereby set up the scene for the
\noemph{coherence} strategy to be presented elsewhere.  We construct an
internal language $\langCartClosed$ for cartesian closed bicategories (where
`ps' stands for \emph{pseudo}), thus reducing the problem of coherence for
cartesian closed bicategories to a property of $\langCartClosed$. This type
theory provides a practical calculus for reasoning in such settings and
directly generalises the STLC~(Simply-Typed Lambda
Calculus)~\cite{Church1940}.  

Our work is motivated by the complexities of calculating in the cartesian
closed bicategories of generalised species~\cite{FioreSpecies} and of
cartesian distributors~\cite{Fiore2015}, specifically for their application
to higher-dimensional category theory~\cite{FioreOpetopicBonn}.  However, the
internal language we present applies to other examples: cartesian closed
bicategories also appear in categorical algebra~\cite{Gambino2017} and game
semantics~\cite{Yamada2018}. 

\subsection{2-dimensional type theories and bicategorical composition} 
\label{sec:2-lambda-calculus}

There is a natural connection between 2-categories and rewriting.  
If objects are types and morphisms are terms, then 2-cells are rewrites
between terms. This idea was explored as early as the 1980s in the work of
Rydeheard \& Stell~\cite{Rydeheard1987} and Power~\cite{Power1989}. For STLC,
Seely~\cite{Seely1987} suggested that \mbox{$\eta$-expansion} and
\mbox{$\beta$-contraction} may naturally be interpreted as the unit and counit
of the adjunction defining (lax) exponentials in a 2-category, an approach
followed by Hilken~\cite{Hilken1996} and advocated by 
Ghani \& Jay~\cite{Jay1995, Ghani1995}. More recently, type-theoretic
constructions modelling 2-categories with strict cartesian closed structure
have been pursued in programming-language theory~\cite{Tabareau2011} and proof
theory~\cite{Hirschowitz2013} while a directed 2-dimensional type theory in
the style of Martin-L\"of~\cite{MartinLof} has been introduced by 
Licata \& Harper~\cite{Licata2011}.

It is crucial to our approach that the equational theory of $\langCartClosed$
does not identify any more structure than the axioms of cartesian closed
bicategories. This entails distinguishing more terms than in STLC. For
instance, note that terms such as
\begin{center} 
$t[u_1 / x_1, u_2 / x_2][v / y]$ 
\:\: \text{and} \:\: 
$t[u_1[v/y] / x_1, u_2[v/y] / x_2]$
\end{center}
are respectively interpreted in a Lambek-style semantics~\cite{Lambek1985} by
the equal maps 
\begin{center}
	$\sem{t} \circ \tup{\sem{u_1}, \sem{u_2}} \circ \sem{v}$ 
  \ and \
	$\sem{t} \circ \tup{\sem{u_1} \circ \sem{v}, \sem{u_2} \circ \sem{v}}$
\end{center}
In contrast, in the \emph{bi}categorical setting these composites are only
isomorphic.\footnote{This issue is similar to that identified by
  Curien~\cite{Curien1993}, who attempts to rectify the mismatch between
  locally cartesian closed categories and \mbox{Martin-L\"of} dependent type
  theory caused by interpreting the (strictly associative) substitution
  operation as a pullback (associative up to coherent isomorphism).} 
Hence, substitution ought to be associative only up to isomorphism. This
places us outside the \mbox{2-categorical} world of previous work, as well as
setting our work apart from type theories in which weak structure is modelled
in a strict language, such as Homotopy Type Theory~(\cf~\cite{hottbook,
Riehl2017}).

\subsection{The type theory $\langCartClosed$}

We will construct the internal language $\langCartClosed$ in stages. First we
will construct the internal language of bicategories
$\langBicat$~(Section~\ref{sec:bicat-type-theory}) and then the internal
language of bicategories with finite products
$\langCart$~(Section~\ref{sec:fp-bicats-type-theory}); the type theory
$\langCartClosed$ extends both these 
systems~(Section~\ref{sec:cc-bicats-type theory}). In each case we construct
the syntactic model and prove an appropriate \mbox{2-dimensional} freeness
universal property.

We introduce substitution formally using a version of 
\Def{explicit substitution}~\cite{Abadi1989, Ritter1997}.
This syntactic structure and the axioms it is subject to are synthesised from
a bicategorification of the \noDef{abstract clones}~\cite{CloneBookRef} of
universal algebra~(Section~\ref{sec:biclones}).  Abstract clones are a natural
bridge between syntactic structure (in the form of intuitionistic 
type theories %~\cite{} 
or 
calculi) %~\cite{}) 
and semantic 
structure (in the form of 
Lawvere theories %~\cite{} 
or cartesian multicategories). %~\cite{Lambek1969}). 

Cartesian closed structure is synthesised from \noemph{universal arrows} at
both the global (2-dimensional) and the local (\mbox{1-dimensional}) levels.
From this approach we recover versions of the usual $\beta\eta$-laws of STLC,
while keeping the rules to a minimum~(Sections~\ref{sec:fp-bicats-type-theory}
and Section~\ref{sec:cc-bicats-type theory}). 
Our formulation points towards similar constructions for
\noDef{tricategories}~(weak \mbox{3-categories}~\cite{Gordon1995, Gurski2013})
or even \noDef{$\infty$-categories}. 

The type theory $\langCartClosed$ is, in a precise sense, a language for
cartesian closed bicategories.  It is capable of being formalised in proof
assistants such as Agda~\cite{agda} and the principled nature of its
construction makes it readily amenable to the addition of further structure.
We leave such extensions for future work.

\section{Bicategories} \label{sec:bicategories} 

We recall the definition of bicategory, pseudofunctor and biequivalence.  For
leisurely introductions
consult~\eg~\cite{Benabou1967},~\cite[\S9]{Street1995}. 

\begin{mydefn}[{\cite{Benabou1967}}] 
A \Def{bicategory} $\baseCat$ consists of 
\begin{itemize} 
\item 
  a class of objects $ob(\baseCat)$, 

\item 
  for every ${X, Y \in ob(\baseCat)}$ a \Def{hom-category} 
  $\big(\baseCat(X, Y), \vert, \id\big)$ with objects \Def{1-cells} 
  ${f : X \to Y}$ and morphisms 
  \mbox{\Def{2-cells}}~\mbox{$\alpha : f \To f' : X \to Y$}; composition of
  2-cells is called \Def{vertical composition}, 

\item 
  for every 
  $X$, $Y$, $Z \in ob(\baseCat)$ %%% HACK !!!
  an \Def{identity} functor ${\Id_X : \catOne \to \baseCat(X,X)}$ and a 
  \Def{horizontal composition} functor
  ${\circ_{X,Y,Z} : \baseCat(Y, Z) \times \baseCat(X, Y) \to \baseCat(X, Z)}$, 

\item 
invertible 2-cells
\begin{align*} 
\a_{h,g,f} : (h \circ g) \circ f &\To h \circ (g \circ f) : W \to Z \\ 
\l_f : \Id_X \circ f &\To f : W \to X \\ 
\r_g : g \circ \Id_X &\To g : X \to Y
\end{align*} 
for every $f : W \to X$, $g : X \to Y$ and $h : Y \to Z$, natural in each of
their arguments and satisfying two coherence laws. %%%~\cite{Benabou1967}.
\end{itemize} 
The functoriality of horizontal composition gives rise to an 
\Def{interchange law}: for suitable 2-cells $\tau, \tau', \sigma, \sigma'$ 
one has 
\mbox{$(\tau' \vert \tau) \circ (\sigma' \vert \sigma) 
  = (\tau' \circ \sigma') \vert (\tau \circ \sigma)$}.  
\end{mydefn}

A morphism of bicategories is called a \noDef{pseudofunctor} (or
\noDef{homomorphism}). %~\cite{Benabou1967}. 
It is a mapping on objects, \mbox{1-cells} and \mbox{2-cells} that preserves
horizontal composition up to isomorphism. 

\begin{mydefn}[{\cite{Benabou1967}}] 
A \Def{pseudofunctor} $F: \baseCat \to \altCat$ between bicategories
$\baseCat$ and $\altCat$ consists of 
\begin{itemize} 
\item 
  a mapping $F : ob(\baseCat) \to ob(\altCat)$, 

\item 
  a functor \mbox{$F_{X,Y} : \baseCat(X,Y) \to \altCat(FX, FY)$} for every
  $X,Y \in ob(\baseCat)$, 

\item 
  an invertible 2-cell \mbox{$\psi_X : \Id_{FX} \To F(\Id_X)$} for every
  \mbox{$X \in ob(\baseCat)$}, 

\item 
  an invertible 2-cell $\phi_{f,g} : F(f) \circ F(g) \To F(f \circ g)$ for
  every $g : X \to Y$ and $f : Y \to Z$, natural in $f$ and $g$ 
\end{itemize} 
subject to three coherence laws. %~\cite{Benabou1967}. 
A pseudofunctor for which $\psi$ and $\phi$ are both the identity is called
\Def{strict}.  
\end{mydefn} 

\begin{myexmp} \label{exmp:bicats} 
Every 2-category is a bicategory and every 2-functor is a strict
pseudofunctor.   
A \mbox{one-object} bicategory is equivalently a monoidal category; a monoidal
functor is equivalently a pseudofunctor between one-object bicategories.
\end{myexmp} 

Bicategorical products and exponentials are defined using the appropriate
notion of adjunction, called a \noDef{biadjunction}.  For our purposes, the
characterisation of biadjoints in terms of 
\noemph{biuniversal arrows}~\cite{Power1998} 
is most natural~(\cf~\cite{FioreSpecies}); this is the bicategorical version of 
the well-known description of adjunctions via universal 
arrows~(\eg~\cite[Chapter III, \S1]{cfwm}).  For biuniversal arrows and their
relationship to biadjunctions, see~\eg~\cite{TFiore2006}. 

\begin{mydefn}[{\cite{Gray1974}}] \label{def:biadjoint} 
Let $F : \baseCat \to \altCat$ be a pseudofunctor. To give a 
\Def{right biadjoint} $U$ to $F$ is to give
\begin{enumerate} 
\item a mapping $U : ob(\altCat) \to ob(\baseCat)$ on objects, 
\item a family of 1-cells $(q_C : FUC \to C)_{C \in ob(\altCat)}$, 
\item for every $B \in ob(\baseCat)$ and \mbox{$C \in ob(\altCat)$} an adjoint equivalence
\begin{small}
\begin{equation*} %\label{diag:Biadjoint}
\begin{tikzcd} 
\baseCat(B, UC) \arrow[bend left = 20]{r}{q_C \circ F(-)} \arrow[phantom]{r}[]{\adjUp} & 
\altCat(FB, C) \arrow[bend left = 20]{l}{(-)^\flat} 
\end{tikzcd}
\end{equation*}
\end{small}
\end{enumerate}
\end{mydefn}

Morphisms of pseudofunctors are called 
\Def{pseudonatural transformations}~\cite{Gray1974} 
and morphisms of pseudonatural transformations are
called~\Def{modifications}~\cite{Benabou1967}.  Bicategories, pseudofunctors,
pseudonatural transformations and modifications organise themselves into a
tricategory we denote $\Bicat$. 

\begin{myexmp} \label{e:ProductBicat} \quad
For every pair of bicategories $\baseCat$ and $\altCat$ there is a bicategory
$\Hom(\baseCat, \altCat)$ of pseudofunctors, pseudonatural transformations and
modifications.
\end{myexmp} 

Bicategories provide a convenient setting for abstractly describing many
categorical concepts~(\eg~\cite{Lawvere1996,Lack2012}). 

\begin{mydefn} \label{def:AdjunctionInBicat} Let $\baseCat$ be a bicategory. 
\begin{enumerate} 
\item 
  An \Def{adjunction} $(A,B, f, g, \eta, \epsilon)$ in $\baseCat$ is a pair of
  objects $(A, B)$ with arrows $f : A \leftrightarrows B : g$ and 2-cells
  \mbox{$\eta : \Id_A \To g \circ f$} and $\epsilon : f \circ g \To \Id_B$
  subject to two \noDef{triangle laws}. 

\item 
  An \Def{equivalence} $(A, B, f, g, \eta, \epsilon)$ in $\baseCat$ is a pair
  of objects $(A,B)$ with arrows $f : A \leftrightarrows B : g$ and invertible
  2-cells $\eta : \Id_A \XRA{\iso} g \circ f$ and 
  $\epsilon : f \circ g \XRA{\iso} \Id_B$. 

\item 
  An \Def{adjoint equivalence} is an adjunction that is also an equivalence.
\end{enumerate} 
\end{mydefn} 

The appropriate notion of equivalence between bicategories is called
\noDef{biequivalence}~\cite{Street1980}. 

\begin{mydefn} 
A \Def{biequivalence} between bicategories $\baseCat$ and $\altCat$ consists
of pseudofunctors $F : \baseCat \leftrightarrows \altCat : G$ with
equivalences $G \circ F \simeq \id_\baseCat$ and 
$F \circ G \simeq \id_\altCat$ in the bicategories $\Hom(\baseCat, \baseCat)$
and $\Hom(\altCat, \altCat)$, respectively.  
\end{mydefn} 

\section{Signatures for 2-dimensional type theories} 

The STLC with constants is determined by a choice of base types and constant
terms~(\eg~\cite{Lambek1985}).  For constants defined in arbitrary contexts
such a choice is determined by a \Def{multigraph}; that is, a set of
\Def{nodes} $A_1, \dots, A_n, B, \dots$ connected by \Def{multiedges}
${\seq{A_1, \dots, A_n} \to B}$.  A multigraph consisting solely of
\Def{edges}~(\ie~multiedges of the form ${\seq A \to B}$) is called a
\Def{graph}.

\begin{mynotation}  \label{not:ind}
In the following definition, and throughout, we write $\ind{A}$ for a finite
sequence $\seq{A_1, \dots, A_n}$ $(n \in \Nat)$.  
\end{mynotation}

\begin{mydefn} \label{def:two-multigraph}
A \Def{2-multigraph} $\graph$ is a set $\nodes{\graph}$ of \Def{nodes}
equipped with a graph 
$\graph(\ind{A}; B)$ of \Def{edges} and \Def{surfaces} for every $n \in \Nat$
and $A_1, \dots, A_n, B \in \nodes{\graph}$. 
A \Def{homomorphism} of 2-multigraphs $h : \graph \to \graph'$ is a map 
$h : \nodes{\graph} \to \nodes{\graph}'$ together with functions
\begin{align*}
h_{A_1, \dots, A_n; B} 
  : \graph(\ind{A}; B) &\to \graph'(\seq{hA_1, \dots, hA_n}\,; hB) 
\\[1mm] 
h_{f,g} 
  : \graph(\ind{A}; B)(f,g) &\to \graph'(\seq{hA_1, \dots, hA_n}\,; hB)(hf,hg)
\end{align*}
for every $n \in \Nat$, $A_1, \dots, A_n, B \in \nodes{\graph}$ and 
$f, g \in \graph(\ind{A}; B)$.  
\end{mydefn}

We denote the category of 2-multigraphs by $\TwoMultiGraph$. The full
subcategory $\TwoGraph$ of \Def{2-graphs} is formed by restricting to
2-multigraphs $\graph$ such that 
$\graph(\seq{A_1, \dots, A_n}\,; B) = \emptyset$ whenever $n \neq 1$.  

\section{Substitution structure up to isomorphism} \label{sec:biclones}

The type theory we shall construct has types, terms, rewrites and an
explicit substitution operation. It is therefore determined by a 2-multigraph
together with a specified substitution structure. Accordingly, to synthesise
our language we introduce an intermediate step between 2-multigraphs and
bicategories, which we call \noDef{biclones}. These are a bicategorification
of the \noDef{abstract clones} of universal algebra~\cite{CloneBookRef}, which
capture a presentation-independent notion of equational theory with
substitution. 

\begin{mydefn} 
An \Def{$S$-biclone} $\clone$ is a set $S$ of \Def{sorts} equipped with the
following for all 
$n$, $m \in \Nat$ %%% HACK !!!
and $X_1, \dots, X_n, Y, Y_1, \dots, Y_m, Z \in S$: 
\begin{itemize} 
\item 
a category $\clone(\ind{X}; Y)$ with objects \Def{1-cells} $f : \ind{X} \to Y$
and morphisms \Def{2-cells} \mbox{$\alpha : f \To g : \ind{X} \to Y$},

\item 
distinguished \Def{projection} functors 
$\p_i : \catOne \to \clone(\ind{X}; X_i)$ for $1\leq i\leq n$, 

\item 
a \Def{substitution} functor 
\begin{gather*}
\mathsf{sub}_{\ind{X}; \ind{Y}; Z}  : \clone(\ind{Y}; Z) \times \prod_{j=1}^m \clone(\ind{X}; Y_j) \to \clone(\ind{X}; Z)
\end{gather*}
which we denote by 
\[
\mathsf{sub}_{\ind{X}; \ind{Y}; Z}\big(f, (g_1, \dots, g_m)\big) := \clonesub{f}{g_1, \dots, g_m}
\]
(we write $\clonesub{t}{\clonesub{\ind{v}}{\ind{w}}}$ or
$\clonesub{t}{\clonesub{v_1}{\ind{w}}, \dots, \clonesub{v_n}{\ind{w}}}$ for
the iterated substitution $\clonesub{t}{\clonesub{v_1}{w_1, \dots, w_l},
\dots, \clonesub{v_n}{w_1, \dots, w_l}}$, \cf~Notation~\ref{not:ind}),

\item 
 natural families of invertible 2-cells 
\begin{align*}
\assoc{t, \ind{u}, \ind{v}} : \clonesub{\clonesub{t}{u_1, \dots, u_n}}{\ind{v}} &\To \clonesub{t}{\clonesub{u_1}{\ind{v}}, \dots, \clonesub{u_n}{\ind{v}}}  \\
\subid{u} : u &\To \clonesub{u}{\p_1, \dots, \p_n} \\
\indproj{k}{u_1, \dots, u_n} : \clonesub{\p_k}{u_1, \dots, u_n} &\To u_k \quad (k = 1,\dots, n)
\end{align*}
for every $t \in \clone(\ind{Y}, Z)$, \mbox{$u_j \in \clone(\ind{X}, Y_j)$},
\mbox{$v_i \in \clone(\ind{W}, X_i)$} and $u \in \clone(\ind{X}, Y)$ ($i = 1,
\dots, n$ and $j = 1, \dots, m$).  
\end{itemize}
This data is subject to two compatibility laws: 
\begin{small}
	\begin{center} 
		\begin{tikzcd}[column sep = small, ampersand replacement = \&] 
		\clonesub{t}{u_1, \dots, u_n} \arrow{r}[yshift=1.5mm]{\clonesub{\subid{}}{u_1, \dots, u_n}} \arrow[swap]{d}{\id} \& 
    \clonesub{\clonesub{t}{\p_1, \dots, \p_n}}{u_1, \dots, u_n}
    \arrow{d}{\assoc{}} \\
		\clonesub{t}{u_1, \dots, u_n} \& 
		\clonesub{t}{\clonesub{\p_1}{u_1, \dots, u_n}, \dots, \clonesub{\p_n}{ u_1, \dots, u_n}} \arrow{l}[yshift=-1.5mm]{\clonesub{t}{\indproj{1}{}, \dots, \indproj{n}{}}} 
		\end{tikzcd} 
	\end{center}
\end{small}
\vspace*{.0125mm}%%% HACK !!!
\begin{small}
	\begin{center}
		\begin{tikzcd}[column sep = small]
		\clonesub{\clonesub{\clonesub{t}{\ind{u}}}{\ind{v}}}{\ind{w}} \arrow[swap]{d}{\assoc{}} \arrow{r}[yshift=1.5mm]{\clonesub{\assoc{}}{\ind{w}}} &
		\clonesub{\clonesub{t}{\clonesub{\ind{u}}{\ind{v}}}}{\ind{w}} \arrow[]{r}[yshift=1.5mm]{\assoc{}} &
		\clonesub{t}{\clonesub{\clonesub{\ind{u}}{\ind{v}}}{\ind{w}}} \arrow{d}{\clonesub{t}{\assoc{}, \dots, \assoc{}}} \\
		\clonesub{\clonesub{t}{\ind{u}}}{\clonesub{\ind{v}}{\ind{w}}} \arrow[swap]{rr}[yshift=-1.5mm]{\assoc{}} &
		\: & 
		\clonesub{t}{\clonesub{\ind{u}}{\clonesub{\ind{v}}{\ind{w}}}}
		\end{tikzcd}			
	\end{center}
\end{small}
When the set $S$ of sorts is clear we refer to an $S$-biclone as simply a
\Def{biclone}.
\end{mydefn} 

Thinking of a bicategory as roughly a 2-category with structure up to
isomorphism, one may think of a biclone as roughly a $\Cat$-enriched clone
with structure up to isomorphism. Indeed, the definition of clone may be
generalised to hold in any cartesian category (and even more
generally,~\eg~\cite{Staton2013, Fiore2017}): if the structural 2-cells
$\subid{}, \proj{}$ and $\assoc{}$ are all the identity, a  biclone is
equivalently a \Def{2-clone}, \ie~a clone in the cartesian category $\Cat$. We
have directed the 2-cells to match the definition of a 
\Def{skew monoidal category}~\cite{Szlachanyi2012}; the definition should
therefore generalise to the lax setting (\cf~also the \Def{lax bicategories}
of Leinster~\cite[\S3.4]{Leinster2004}).

Every $S$-biclone~$\clone$ has an underlying \Def{linear-core}
bicategory~$\lincore\clone$ with objects~$S$ and
hom-categories~$\lincore\clone(X,Y)
  =\clone\big(\seq X\,;Y\big)$~(\cf~\cite{Hermida2000}).
Every \mbox{2-multigraph} freely induces a sorted biclone, and one may
introduce bicategorical substitution structure into a type theory with base
types, constant terms and constant rewrites specified by a
\mbox{2-graph}~$\graph$ by postulating the structure of the free sorted
biclone on $\graph$ and then restricting it to its underlying linear core.
This is the gist of the following section.

\begin{figure*}[!ht]
\centering
{\small
\begin{minipage}{\textwidth}
\begin{mdframed}
\centering
\unaryRule	{}
			{x_1 : A_1, \dots, x_n : A_n \vdash x_k : A_k}
			{var $(1 \leq k \leq n)$}
\unaryRule	{\big(c \in \graph(A_1, \dots, A_n; B)\big)}
			{x_1 : A_1, \dots, x_n : A_n \vdash c(x_1, \dots, x_n) : B}
			{const}
		
\binaryRule	{x_1 : A_1, \dots, x_n : A_n \vdash t : B}
			{(\Delta \vdash u_i : A_i)_{i= 1, \dots, n}}
			{\Delta \vdash \hcomp{t}{x_1 \mapsto u_1, \dots, x_n \mapsto u_n} : B}
			{horiz-comp}
\caption{Introduction rules on basic terms \label{r:basic-terms}}
\end{mdframed}
\end{minipage}

\begin{minipage}{\textwidth}
\begin{mdframed}
\centering
\begin{prooftree}
\AxiomC{$x_1 : A_1, \dots, x_n : A_n \vdash t : B$}
\RightLabel{\scriptsize $\subid{}$-intro}
\UnaryInfC{$x_1 : A_1, \dots, x_n : A_n \vdash \subid{t} : \birewrite{t}{\hcomp{t}{x_i \mapsto x_i}} : B$}
\noLine
\UnaryInfC{$x_1 : A_1, \dots, x_n : A_n \vdash \subid{t}^{-1} : \birewrite{\hcomp{t}{x_i \mapsto x_i}}{t} : B$}
\end{prooftree}\vspace{0.5\treeskip}

\begin{prooftree}
\AxiomC{$x_1 : A_1, \dots, x_n : A_n \vdash x_k : A_k$}
\AxiomC{$(\Delta \vdash u_i : A_i)_{i = 1, \dots, n}$}
\RightLabel{\scriptsize $\indproj{k}{}$-intro $(1 \leq k \leq n)$}
\BinaryInfC{$\Delta \vdash \indproj{k}{u_1, \dots, u_n} : \birewrite{\hcomp{x_k}{x_i \mapsto u_i}}{u_k} : A_k$}
\noLine
\UnaryInfC{$\Delta \vdash \indproj{-k}{u_1, \dots, u_n} :  \birewrite{u_k}{\hcomp{x_k}{x_i \mapsto u_i}} : A_k$}
\end{prooftree}\vspace{0.5\treeskip}

\begin{prooftree}
\AxiomC{$y_1 : B_1, \dots, y_n : B_n \vdash t : C$}
\AxiomC{$(x_1 : A_1, \dots, x_m : A_m \vdash v_i : B_i)_{i = 1, \dots, n}$}
\AxiomC{$(\Delta \vdash u_j : A_j)_{j = 1, \dots m}$}
\RightLabel{\scriptsize $\assoc{}$-intro}
\TrinaryInfC{$\Delta \vdash \assoc{t, \ind{v}, \ind{u}} : \rewrite{\hcomp{\hcomp{t}{y_i \mapsto v_i}}{x_j \mapsto u_j}}{\hcomp{t}{y_i \mapsto \hcomp{v_i}{x_j \mapsto u_j}}} : C$}
\noLine
\UnaryInfC{$\Delta \vdash \assoc{t, \ind{v}, \ind{u}}^{-1} : \rewrite{\hcomp{t}{y_i \mapsto \hcomp{v_i}{x_j \mapsto u_j}}}{\hcomp{\hcomp{t}{y_i \mapsto v_i}}{x_j \mapsto u_j}} : C$}
\end{prooftree}
\caption{Introduction rules on structural rewrites \label{r:structural-rewrites}}
\end{mdframed}
\end{minipage}

\begin{minipage}{\textwidth}
\begin{mdframed}
\centering
\unaryRule	{\Gamma \vdash t : A}
			{\Gamma \vdash \id_t : \birewrite{t}{t} : A}
			{$\id$-intro}
\unaryRule		{\big(\sigma \in \graph(A_1, \dots, A_n; B)(c, c')\big)}
				{x_1: A_1, \dots, x_n : A_n \vdash \sigma(x_1, \dots, x_n) : \rewrite{c(x_1, \dots, x_n)}{c'(x_1, \dots, x_n)} : B}
				{2-const}
\vspace{-\treeskip}
\caption{Introduction rules on basic rewrites \label{r:basic-rewrites}}
\end{mdframed}
\end{minipage}

\begin{minipage}{\textwidth}
\begin{mdframed}
\centering
\binaryRule	{\Gamma \vdash \tau' : \rewrite{t'}{t''} : A}
			{\Gamma \vdash \tau : \rewrite{t}{t'} : A}
			{\Gamma \vdash \tau' \vertsub{t, t',t''} \tau : \rewrite{t}{t''} : A}
			{vert-comp}
\binaryRule	{x_1 : A_1, \dots, x_n : A_n \vdash \tau : \rewrite{t}{t'} : B}
			{(\Delta \vdash \sigma_i : \rewrite{u_i}{u'_i} : A_i)_{i = 1, \dots, n}}
			{\Delta \vdash \horizComp{\tau}{x_i \mapsto \sigma_i} : \rewrite{\hcomp{t}{x_i \mapsto u_i}}{\hcomp{t'}{x_i \mapsto u_i'}} : B}
			{horiz-comp}
\caption{Constructors on rewrites \label{r:constructors-on-rewrites}}
\end{mdframed}
\end{minipage}
}

\begin{manyfigcap}
Introduction rules for $\langBicat$, $\langCart$ and $\langCartClosed$.  For $\langBicat$ these rules are restricted to \emph{unary} contexts.
\end{manyfigcap}
\end{figure*}

\section{A type theory for bicategories}  \label{sec:bicat-type-theory}

We follow the tradition of 2-dimensional type theories consisting of types,
terms and
rewrites,~\eg~\cite{Seely1987,Hilken1996,Licata2011,Hirschowitz2013}.
Following Hilken~\cite{Hilken1996}, we consider two forms of judgement.
Alongside the usual $\Gamma \vdash t : A$ to indicate `term $t$ has type $A$
in context $\Gamma$', we write $\Gamma \vdash \tau : \rewrite{t}{t'} : A$ to
indicate `$\tau$ is a rewrite from term $t$ of type $A$ to term $t'$ of type
$A$, in context~$\Gamma$'. 

For now we do not want to assume our model has products so we restrict to
unary contexts. Base types, constants and rewrites are therefore specified by
a 2-graph $\graph$. The term introduction rules are collected in
Figure~\ref{r:basic-terms} and the rewrite introduction rules in
Figures~\ref{r:structural-rewrites}--\ref{r:constructors-on-rewrites}. We
denote the language thus defined by $\langBicat(\graph)$.

The terms are variables~$x$, $y, \dots$, %%% HACK !!!
constant terms~$c(x)$, $c'(x), \dots$ %%% HACK !!!
and \Def{explicit substitutions} $\hcomp{t}{x \mapsto u} $. 
Thus for every term $u$ and term $t$ with free variable $x$ we postulate a
term $\hcomp{t}{x \mapsto u}$; this is a formal analogue of the term~$t[u/x]$
defined by the meta-operation of capture-avoiding
substitution~(\cf~\cite{Abadi1989, Ritter1997}). The variable $x$ is bound by
this operation, and we work with terms up to $\alpha$-equivalence defined in
the standard way. 
Instead of being typed in the empty context, constants are given by the edges
of $\graph$.  The restriction to unary contexts ensures the syntactic model is
a bicategory, rather than a biclone.  When we construct the type theory for
bicategories with finite products we shall allow constants and explicit
substitution to be multi-ary.

The grammar for rewrites is synthesised from the free biclone on $\graph$.
The \Def{structural rewrites} $\assoc{}, \subid{}$ and $\proj{}$ witness the
three laws of a biclone;  we slightly abuse notation by simultaneously
introducing these rewrites and their inverses. The \Def{constant rewrites}
$\sigma(x)$ are specified by the surfaces of $\graph$ and for every term $t$
we have an \Def{identity rewrite} $\id_t$. The explicit substitution operation
mirrors that for terms while vertical composition is captured by a binary
operation on rewrites~(\cf~\cite{Hilken1996, Hirschowitz2013}). 
We make use of the same conventions on binding and $\alpha$-equivalence as for
terms. 

\begin{mynotation} \label{not:BicatConventions}
We adopt the following abuses of notation.
\begin{enumerate}
\item 
  Writing $t$ for $\id_t$ in a rewrite. 
\item 
  Writing $\hcomp{t}{x_i \mapsto u_i}_{i=1}^n$ or simply 
  $\hcomp{t}{x_i \mapsto u_i}$ for 
  $\hcomp{t}{x_1 \mapsto u_1, \dots, x_n \mapsto u_n}$, and similarly on
  rewrites. 
\item \label{c:constants} 
  Writing $\hcomp{c}{t_1, \dots, t_n}$ for the explicit substitution
  $\hcomp{c(x_1, \dots, x_n)}{x_i \mapsto t_i}$ for constants $c$, and
  similarly on rewrites.
\end{enumerate}
\end{mynotation}

The equational theory $\equiv$ on rewrites is derived directly from the axioms
of a biclone; these are collected in
\mbox{Figures~\ref{r:hom-categories-axioms}--\ref{r:biclone-laws}}. 
In this extended abstract, the rules for the invertibility of the structural
rewrites and the congruence rules on~$\equiv$ are omitted.

\label{sec:PropertiesOfLangBicat}
The type theory $\langBicat(\graph)$ satisfies the expected well-formedness
properties (\cf~\cite[Chapter 4]{Crole1994} for STLC). 
Uniqueness of typing is obtained by adding type annotations on bound
variables, constants and vertical compositions; we omit this extra information
for readability.

\subsection{The syntactic model for $\langBicat$}
\label{sec:bicat:syntactic-model}

We construct the syntactic model for $\langBicat(\graph)$ and prove that it
enjoys a 2-dimensional freeness universal property analogous to that
of~\cite[\S2.2.1]{Gurski2013}.  Put colloquially, $\langBicat(\graph)$ is the
\noDef{internal language} for bicategories with signature a 2-graph~$\graph$.
The bicategorical structure is induced directly from the biclone structure.

\begin{myconstr} \label{constr:bicat-termcat} 
For any 2-graph $\graph$, define a bicategory $\termCat(\graph)$ as follows.
Objects are unary contexts $(x : A)$. The hom-category
\mbox{$\termCat(\graph)\big((x:A),(y:B)\big)$} has objects
$\alpha$-equivalence classes of derivable terms 
\mbox{$(x : A \vdash t :B)$} and morphisms $\alpha{\equiv}$-equivalence
classes of rewrites \mbox{$(x : A \vdash \tau : \rewrite{t}{t'} : B)$} under
vertical composition. Horizontal composition is given by explicit substitution
and the identity on $(x : A)$ by the \rulename{var} rule $( x: A \vdash x :
A)$. The structural isomorphisms $\l, \r$ and $\a$ are $\proj{}$,
$\subid{}^{-1}$ and $\assoc{}$, respectively.
\end{myconstr} 

\begin{myconstr} \label{constr:DefnOfBicatExtension}  
For any 2-graph $\graph$, bicategory $\baseCat$ and \mbox{2-graph}
homomorphism $h : \graph \to \baseCat$, the semantics of 
Figure~\ref{fig:extension-pseudofunctor} restricted to $\termCat(\graph)$
induces a strict pseudo\-functor $h\semext : \termCat(\graph) \to \baseCat$.
\end{myconstr} 

The universal \emph{extension} 
pseudofunctor~$h\semext$ cannot be characterised by a strict universal
property; for instance, for each type~$A$ the bicategory~$\termCat(\graph)$
contains countably many equivalent objects~${(x:A)}$ for $x$ ranging over
variables. To obtain the desired universal property, we restrict to a
sub-bicategory in which there is a single variable name.

\begin{myconstr} \label{constr:bicat-termcat-fixedvar} 
Let $\termCatSlim(\graph)$ denote the bicategory with objects unary contexts
$(x : A)$ for $x$ a \emph{fixed} variable and horizontal and vertical
composition operations as in $\termCat(\graph)$.  
\end{myconstr}

The bicategories $\termCatSlim(\graph)$ and $\termCat(\graph)$ are
biequivalent, and the restricted model $\termCatSlim(\graph)$ is free on
$\graph$ in the following sense. 

\begin{mythm} \label{prop:BicatTermModelUMP} 
Let $\graph$ be a 2-graph. 
For any bicategory~$\baseCat$ and 2-graph 
homomorphism $h : \graph \to \baseCat$,
there exists a unique strict pseudofunctor 
$h\semext : \termCatSlim(\graph) \to \baseCat$ such that
$h\semext \circ \inc = h$, for 
$\inc : \graph \hookrightarrow \termCatSlim(\graph)$ the inclusion.  
\end{mythm} 

The full model $\termCat(\graph)$ therefore satisfies the universal
property up to biequivalence.  Proving
the freeness universal property in this way has the benefit of allowing us to
establish uniqueness without reasoning about uniqueness of pseudonatural
transformations or modifications.
It follows that $\langBicat$ is an internal language for bicategories.  By
Example~\ref{exmp:bicats}, restricting to a single base type gives rise to an
internal language for monoidal categories.

The argument in this section applies with only minor adjustments to biclones.
Thus, allowing $\langBicat(\graph)$ to have contexts with fixed variables 
over a 2-multigraph $\graph$ gives rise to the free biclone on $\graph$ in the
sense of 
Theorem~\ref{prop:BicatTermModelUMP}. 
This relies on a straightforward generalisation of the notions of
pseudofunctor, transformation and modification.  
Details will be given elsewhere.

\begin{figure*}[!h]

\centering
{\small

\begin{minipage}{\textwidth}
\begin{mdframed}
\centering
\unaryRule	{\Gamma \vdash \tau : \rewrite{t}{t'} : A}
			{\Gamma \vdash \tau \vert \id_t \equiv \tau : \rewrite{t}{t'} : A}
			{$\vert$-right-unit}
\unaryRule	{\Gamma \vdash \tau : \rewrite{t}{t'} : A}
{\Gamma \vdash \tau \equiv \id_{t'} \vert \tau : \rewrite{t}{t'} : A}
{$\vert$-left-unit}

\trinaryRule	{\Gamma \vdash \tau'' : \rewrite{t''}{t'''} : A}
				{\Gamma \vdash \tau' : \rewrite{t'}{t''} : A}
				{\Gamma \vdash \tau : \rewrite{t}{t'} : A}
				{\Gamma \vdash (\tau'' \vert \tau') \vert \tau \equiv \tau'' \vert (\tau' \vert \tau) : \rewrite{t}{t'''} : A}
				{$\vert$-assoc}
\caption{Categorical structure of vertical composition \label{r:hom-categories-axioms}}
\end{mdframed}
\end{minipage}

\begin{minipage}{\textwidth}
\begin{mdframed}
\centering
\binaryRule	{x_1 : A_1, \dots, x_n : A_n \vdash t : B}
			{(\Delta \vdash u_i : A_i)_{i = 1, \dots, n}}
			{\Delta \vdash \hcomp{\id_t}{x_i \mapsto u_i} \equiv \id_{\hcomp{t}{x_i \mapsto u_i}} : \rewrite{\hcomp{t}{x_i \mapsto u_i}}{\hcomp{t}{x_i \mapsto u_i}} : B}
			{$\id$-preservation}

\begin{small}
\begin{prooftree}
\AxiomC{$x_1 : A_1, \dots, x_n : A_n \vdash \tau : \rewrite{t}{t'} : B$}
\noLine
\UnaryInfC{$x_1 : A_1, \dots, x_n : A_n \vdash \tau' : \rewrite{t'}{t''} : B$}
\AxiomC{$(\Delta \vdash \sigma_i : \rewrite{u_i}{u_i'} : A_i)_{i=1,\dots, n}$}
\noLine
\UnaryInfC{$(\Delta \vdash \sigma_i' : \rewrite{u_i'}{u_i''} : A_i)_{i=1,\dots, n}$}

\RightLabel{{\scriptsize interchange}}
\BinaryInfC{$\Delta \vdash \horizComp{\tau'}{x_i \mapsto \sigma_i'} \vert \horizComp{\tau}{x_i \mapsto \sigma_i} \equiv \horizComp{(\tau' \vert \tau)}{x_i \mapsto \sigma_i' \vert \sigma_i} : \rewrite{\hcomp{t}{x_i \mapsto u_i}}{\hcomp{t''}{x_i \mapsto u_i''}} : B$}
\end{prooftree}
\end{small}
\caption{Preservation rules \label{r:hcomp-functor}}
\end{mdframed}
\end{minipage}

\begin{minipage}{\textwidth}
\begin{mdframed}
\centering
\unaryRule	{(\Delta \vdash \sigma_i : \rewrite{u_i}{u_i'} : A_i)_{i = 1, \dots, n}}
			{\Delta \vdash \indproj{k}{u_1', \dots, u_n'} \vert \hcomp{x_k}{x_i \mapsto \sigma_i} \equiv \sigma_k \vert \indproj{k}{u_1, \dots, u_n} : \rewrite{\hcomp{x_k}{x_i \mapsto u_i}}{u_k'} : A_k}
			{$(1 \leq k \leq n)$}

\unaryRule	{x_1 : A_1, \dots, x_n : A_n \vdash \tau : \rewrite{t}{t'} : B}
			{x_1 : A_1, \dots, x_n : A_n \vdash \subid{t'} \vert \tau \equiv \hcomp{\tau}{x_i \mapsto x_i} \vert \subid{t} : \rewrite{t}{\hcomp{t'}{x_i \mapsto x_i}} : B}
			{}
			
\begin{footnotesize}
\begin{prooftree}
\alwaysNoLine
\AxiomC{$y_1 : B_1, \dots, y_n : B_n \vdash \tau : \rewrite{t}{t'} : C$}
\AxiomC{$(x_1 : A_1, \dots, x_m : A_m \vdash \sigma_i : \rewrite{v_i}{v_i'} : B_i)_{i = 1, \dots, n}$}
\AxiomC{$(\Delta \vdash \mu_j : \rewrite{u_j}{u_j'} : A_j)_{j = 1, \dots m}$}
\alwaysSingleLine
\TrinaryInfC{$\Delta \vdash \assoc{t', \ind{v}, \ind{u}} \vert \hcomp{\hcomp{\tau}{y_i \mapsto \sigma_i}}{x_j \mapsto \mu_j} \equiv \hcomp{\tau}{y_i \mapsto \hcomp{\sigma_i}{x_j \mapsto \mu_j}} \vert  \assoc{t, \ind{v}, \ind{u}} : \rewrite{\hcomp{\hcomp{t}{y_i \mapsto v_i}}{x_j \mapsto u_j}}{\hcomp{t'}{y_i \mapsto \hcomp{v_i'}{x_j \mapsto u_j'}}} : C$}
\end{prooftree} 
\end{footnotesize}
\caption{Naturality rules on structural rewrites \label{r:structural-rewrites-nat}}
\end{mdframed}
\end{minipage}

\begin{minipage}{\textwidth}
\begin{mdframed}
\centering
\begin{prooftree}
\AxiomC{$x_1 : A_1, \dots, x_n : A_n \vdash t : B$}
\AxiomC{$(\Delta \vdash u_i : A_i)_{i = 1, \dots, n}$}
\BinaryInfC{$\Delta \vdash \hcomp{t}{x_i \mapsto \indproj{i}{\ind{u}}} \vert \assoc{t,\ind{x},\ind{u}} \vert \hcomp{\subid{t}}{x_i \mapsto u_i} \equiv \id_{\hcomp{t}{x_i \mapsto u_i}} : \rewrite{\hcomp{t}{x_i \mapsto u_i}}{\hcomp{t}{x_i \mapsto u_i}} : B$}
\end{prooftree} \vspace{\treeskip}

\begin{footnotesize}
\begin{prooftree}
\alwaysNoLine
\AxiomC{$z_1 : C_1, \dots, z_l : C_l \vdash t : D$}
\UnaryInfC{$(y_1 : B_1, \dots, y_n : B_n \vdash w_j : C_k)_{k = 1, \dots, l}$}
\AxiomC{$(x_1 : A_1, \dots, x_m : A_m \vdash v_i : B_i)_{i = 1, \dots, n}$}
\UnaryInfC{$(\Delta \vdash u_j : A_j)_{j = 1, \dots m}$}
\alwaysSingleLine
\BinaryInfC{$\Delta \vdash \hcomp{t}{z_k \mapsto \assoc{w_k, \ind{v}, \ind{u}}} \vert \assoc{t, \hcomp{\ind{w}}{y_j \mapsto v_j}, \ind{u}} \vert \hcomp{\assoc{t, \ind{w}, \ind{v}}}{x_j \mapsto u_j} \equiv \assoc{t, \ind{w}, \hcomp{\ind{v}}{x_j \mapsto u_i}} \vert \assoc{\hcomp{t}{z_k \mapsto w_k}, \ind{v}, \ind{u}}$\hspace{9mm}}
\noLine
\UnaryInfC{\hspace{70mm}$: \rewrite{\hcomp{\hcomp{\hcomp{t}{z_k \mapsto w_k}}{y_i \mapsto v_i}}{x_j \mapsto u_j}}{\hcomp{t}{z_k \mapsto \hcomp{w_k}{y_i \mapsto \hcomp{v_i}{x_j \mapsto u_j}}}} : D$}
\end{prooftree}
\end{footnotesize} 

\caption{Biclone laws \label{r:biclone-laws}}
\end{mdframed}
\end{minipage}
}

\begin{manyfigcap}
Equational theory for structural rewrites in $\langBicat$, $\langCart$ and $\langCartClosed$. For $\langBicat$ the rules are restricted to unary contexts.
\end{manyfigcap}
\end{figure*}

\section{fp-Bicategories} \label{sec:fp-bicats}

We recall the notion of bicategory with finite products, defined as
a~\noDef{bilimit}~\cite{Gray1974}. To avoid confusion with the `cartesian
bicategories' of Carboni~and~Walters~\cite{Carboni1987,Carboni2008}, we use
the term \Def{fp-bicategories}.  fp-Bicategories are the objects of a
tricategory~$\CartBicat$. 

It is convenient to work directly with $n$-ary products~${(n \in \Nat)}$.
This reduces the need to deal with the equivalent objects given
by rebracketing binary products. For 
bicategories~$\baseCat_1, \dots, \baseCat_n$ the \Def{product bicategory}
$\prod_{i=1}^n \baseCat_i$ has objects~$(B_1, \dots, B_n) \in \prod_{i=1}^n
ob(\baseCat_i)$ and structure
given pointwise.  An \Def{fp-bicategory} is therefore a bicategory~$\baseCat$
equipped with a right biadjoint to the diagonal 
pseudofunctor ${\Delta_n : \baseCat \to \baseCat^{{\times} n} 
  : B \mapsto (B, \dots, B)}$ 
for each $n \in \Nat$. This unwinds to the following definition. 

\begin{mydefn}  \label{def:fp-bicat}
An \Def{fp-bicategory} is a bicategory $\baseCat$ equipped with the following
data for every $n \in \Nat$ and \mbox{$A_1, \dots, A_n \in \baseCat$}:
\begin{enumerate} 
\item \label{c:fp-bicat:obj} 
  a \Def{product} object $\prodop_n(A_1, \dots, A_n)$, 

\item \label{c:fp-bicat:proj} 
  \Def{projection} 1-cells $\pi_k : \prodop_n(A_1, \dots, A_n) \to A_k$ for 
  ${1 \leq k \leq n}$, 

\item \label{c:fp-bicat:adj-equiv} 
  for every $X \in \baseCat$ an adjoint equivalence
\begin{equation} \label{diag:fp-bicategory}
\begin{small}
\begin{tikzcd}
	\baseCat\big(X, \prodop_n (A_1, \dots,A_n)\big) \arrow[bend left = 20]{r}{(\pi_1 \circ -, \dots, \pi_n \circ -)} \arrow[phantom]{r}[xshift=-0.3em]{\adjUp} &
	\prod_{i=1}^n \baseCat(X, A_i) \arrow[bend left = 20]{l}{\tup{-, \dots, =}}
\end{tikzcd}
\end{small}
\end{equation}
\end{enumerate}
We call the right adjoint $\tup{-, \dots, =}$ the \Def{$n$-ary tupling}.
\end{mydefn} 

\begin{myremark}
One obtains a \Def{lax} $n$-ary product structure by merely asking for an
adjunction in diagram~(\ref{diag:fp-bicategory}).  
\end{myremark}

\begin{myexmp} \label{ex:CartesianBicatABiclone} 
Every small fp-bicategory $\big(\baseCat, \Pi_n(-)\big)$ defines an
$ob(\baseCat)$-biclone $\cloneinto\baseCat$ by setting
$\cloneinto\baseCat(\seq{X_1, \dots, X_n}\,; Y)$ to be 
$\baseCat\big(\prodop_n(X_1, \dots, X_n), Y\big)$.
\end{myexmp} 

\begin{mynotation}  \label{not:products} \mbox{}
\begin{enumerate}
\item \label{c:types} 
  We write $A_1 \times \dots \times A_n$ or $\prod_{i=1}^n A_i$ for $\prod_n
  (A_1, \dots, A_n)$ and denote the terminal object $\prod_0 ()$ by
  $\unittype$.
\item \label{c:maps} 
  We write $\tup{f_i}_{i=1, \dots, n}$ or simply $\tup{\ind{f}}$ for the
  $n$-ary tupling $\tup{f_1, \dots, f_n}$.
\end{enumerate}
\end{mynotation} 

There are different ways of specifying the adjoint
equivalence~(\ref{diag:fp-bicategory}) (see~\eg~\cite[Chapter~IV]{cfwm}). One
option is to specify an invertible unit and counit subject to naturality and
triangle laws.  This matches the \mbox{$\eta$-expansion} and
\mbox{$\beta$-reduction} rules of
STLC~(\cf~\cite{Seely1987,Jay1995,Ghani1995,Hirschowitz2013}), but in the
pseudo or lax settings requires a cumbersome proliferation of introduction
rules.  Instead, we characterise the counit
$\epsilonTimesname=(\epsilonTimesInd{1}{}, \dots, \epsilonTimesInd{n}{})$ as a
universal arrow.  Thus, for any finite family of 1-cells 
$(t_i : X \to A_i)_{i=1, \dots, n}$ we
require a 1-cell $\tup{t_1, \dots, t_n} : X \to \prod_n(A_1, \dots, A_n)$ and
a family of 2-cells \mbox{$(\epsilonTimesInd{k}{t_1, \dots, t_n} : \pi_k \circ
  \tup{\ind{t}} \To t_k)_{k = 1, \dots, n}$}, universal in the sense that for
any family of 2-cells \mbox{$(\alpha_i : \pi_i \circ u \To t_i : \Gamma \to
  A_i)_{i = 1, \dots, n}$} there exists a unique 2-cell
\mbox{$\transTimes{\alpha_1, \dots, \alpha_n} : u \To \tup{\ind{t}} : \Gamma
  \to \prod_{i=1}^n A_i$} such that 
\begin{center}
$\epsilonTimesInd{k}{t_1, \dots, t_n} \vert \big(\pi_k \circ
\transTimes{\alpha_1, \dots, \alpha_n}\big) = \alpha_k : \pi_k \circ u \To
t_k$ 
\end{center}
for $k = 1, \dots, n$.

\begin{myexmp} 
In any bicategory, unary-product structure may be chosen to be canonically
given as follows: $\Pi_1(A)=A$, $\pi_1^A=\Id_A$, $\tup t = t$ and
$\varpi_t=\l_t:\Id\circ t\To t$.
\end{myexmp} 

\begin{myremark} \label{rem:nAryProductsDeterminedUpToEquiv} 
As it is well-known, product structure is unique up to equivalence.  Given
adjoint equivalences ${(g : C \leftrightarrows \Pi_{i=1}^n B_i : h)}$
and ${(e_i : B_i \leftrightarrows A_i : f_i)_{i=1, \dots, n}}$
in a bicategory $\baseCat$, the composite 
\begin{equation*} 
{\scriptsize
\begin{tikzcd} 
& 
\baseCat(X, \prod_{i=1}^n B_i) 
\arrow[bend left = 20]{r}{(\pi_1 \circ -, \dots, \pi_n \circ -)} 
\arrow[bend left = 20]{dl}{h\circ-}
\arrow[phantom]{r}[xshift=0em]{\adjUp} 
& 
\prod_{i=1}^n \baseCat(X, B_i)
\arrow[bend left = 20]{l}{\tup{-, \dots, =}} 
\arrow[bend left = 20]{dr}{\Pi_{i=1}^n(e_i \circ -)} 
\arrow[phantom]{rd}[xshift=-.5em]{\adjUp} 
&
\\
\arrow[phantom]{ru}[xshift=0em,yshift=0em]{\adjUp} 
\baseCat(X, C)
\arrow[bend left = 20]{ur}{g\circ-}
& 
& 
& 
\prod_{i=1}^n \baseCat(X, A_i) 
\arrow[bend left = 20]{ul}{\Pi_{i=1}^n(f_i \circ -)} 
\end{tikzcd}
} 
\end{equation*} 
yields an adjoint equivalence
\begin{equation*} 
\begin{tikzcd} 
\baseCat(X, C)
\arrow[bend left = 20]{rr}
{(\,
  ((e_1\circ\pi_1)\circ g)\circ- , \dots , ((e_n\circ \pi_n)\circ g)\circ-
  \,)} 
\arrow[phantom]{rr}[xshift=0em]{\adjUp} 
&& 
\prod_{i=1}^n \baseCat(X, A_i) 
\arrow[bend left = 20]{ll}{h\circ\tup{f_1\circ- , \dots , f_n\circ=}} 
\end{tikzcd}
\end{equation*} 
presenting $C$ as the product of $A_1,\ldots,A_n$.
\end{myremark} 

\begin{mydefn} 
An \Def{fp-pseudofunctor} $(F, \prodPres)$ between \mbox{fp-bicategories}
$\big(\baseCat, \Pi_n(-)\big)$ and $\big(\altCat, \Pi_n(-)\big)$ is a
pseudofunctor $F : \baseCat \to \altCat$ equipped with adjoint equivalences
\[\textstyle
  \big( 
   F\big(\prod_{i=1}^n A_i\big) , 
   \prod_{i=1}^n F(A_i) , 
   \tup{F\pi_1, \dots, F\pi_n} , 
   \prodPres_{A_1, \dots, A_n} %, 
   %\eta^\times , 
   %\varepsilon^\times
\big)\] 
for every $n \in \Nat$ and $A_1, \dots, A_n \in \baseCat$. 

We call $(F, \prodPres)$ \Def{strict} if $F$ is strict and satisfies
\begin{align*}
F\big(\Pi_n(A_1,\ldots,A_n)\big) &= \Pi_n(FA_1,\ldots,FA_n) 
\\
F(\pi_i^{A_1,\ldots,A_n}) &= \pi_i^{FA_1,\ldots,FA_n} 
\\
F\tup{t_1,\ldots,t_n} &= \tup{Ft_1,\ldots,Ft_n} 
\\
F\varpi^{(i)}_{t_1,\ldots,t_n} &= \varpi^{(i)}_{Ft_1,\ldots,Ft_n} 
\\
\prodPres_{A_1,\ldots,A_n} &= \Id_{\Pi_n(FA_1,\ldots,FA_n)}
\end{align*}
and the adjoint equivalences are canonically induced by the %canonical
$2$-cells 
$\transTimes{\r_{\pi_1},\ldots,\r_{\pi_n}}
 : \Id\XRA\iso\tup{\pi_1,\ldots,\pi_n}$.
\end{mydefn} 
Thus, a strict fp-pseudofunctor strictly preserves both (global) biuniversal
arrows and (local) universal arrows.

\begin{figure*}[!h]
{\small
\begin{minipage}{\textwidth}
\begin{mdframed}
\centering
\unaryRule	{\Gamma \vdash t_1 : A_1 \quad \dots \quad \Gamma \vdash t_n : A_n}
			{\Gamma \vdash \pair{t_1, \dots, t_n} : \prodop_n (A_1, \dots, A_n)}
			{$n$-pair}
\unaryRule	{\phantom{a}}
			{p : \prodop_n(A_1, \dots, A_n) \vdash \pi_k(p) : A_k}
			{$k$-proj ($1 \leq k \leq n$)}
\vspace{-\treeskip}
\caption{\label{r:cart:products-terms} Terms for product structure}
\end{mdframed}
\end{minipage}

\begin{minipage}{\textwidth}
\begin{mdframed}
\centering
\unaryRule	{\Gamma \vdash t_1 : A_1 \qquad \dots \qquad \Gamma \vdash t_n : A_n}
			{\Gamma \vdash \epsilonTimesInd{k}{t_1, \dots, t_n} : \rewrite{\hcomp{\pi_k}{\pair{t_1, \dots, t_n}}}{t_k} : A_k} 
			{$\epsilonTimesInd{k}{}$-intro ($1 \leq k \leq n$)}
\binaryRule	{\Gamma \vdash u : \prodop_n(A_1, \dots, A_n)}
			{\Gamma \vdash \alpha_1 : \rewrite{\hcomp{\pi_1}{u}}{t_1} : A_1	\quad \dots \quad \Gamma \vdash \alpha_n : \rewrite{\hcomp{\pi_n}{u}}{t_n} : A_n}
			{\Gamma \vdash \transTimes{\alpha_1, \dots, \alpha_n} : \rewrite{u}{\pair{t_1, \dots, t_n}} : \prodop_n(A_1, \dots, A_n)}
			{$\transTimesSymb$-intro}
\caption{Rewrites for product structure \label{r:cart:ProductsRewrites}}
\end{mdframed}
\end{minipage}

\begin{minipage}{\textwidth}
\begin{mdframed}
\centering
\unaryRule	{\Gamma \vdash \alpha_1 : \rewrite{\hcomp{\pi_1}{u}}{t_1} : A_1 \quad \dots \quad \Gamma \vdash \alpha_n : \rewrite{\hcomp{\pi_n}{u}}{t_n} : A_n}
			{\Gamma \vdash \alpha_k \equiv \epsilonTimesInd{k}{t_1, \dots, t_n} \vert \hcomp{\pi_k}{\transTimes{\alpha_1, \dots, \alpha_n}} :\rewrite{\hcomp{\pi_k}{u}}{t_k} : A_k}
			{U1 ($1 \leq k \leq n$)}
			
\unaryRule	{\Gamma \vdash \gamma : \rewrite{u}{\pair{t_1, \dots, t_n}} : \prodop_n(A_1, \dots, A_n)}
			{\Gamma \vdash \gamma \equiv \transTimes{\epsilonTimesInd{1}{t_1, \dots, t_n} \vert \hcomp{\pi_1}{\gamma}, \dots, \epsilonTimesInd{n}{t_1, \dots, t_n} \vert \hcomp{\pi_n}{\gamma}} : \rewrite{u}{\pair{t_1, \dots, t_n}} : \prodop_n(A_1, \dots, A_n)}
			{U2}
\unaryRule{\big(\Gamma \vdash \alpha_i \equiv  \alpha'_i : \rewrite{\hcomp{\pi_i}{u}}{t_i} : A_i\big)_{i = 1, \dots, n} }
		{\Gamma \vdash \transTimes{\alpha_1, \dots, \alpha_n} \equiv \transTimes{\alpha'_1, \dots, \alpha'_n} : \rewrite{u}{\pair{t_1, \dots, t_n}} : \prodop_n(A_1, \dots, A_n)}
			{cong}
\caption{Universal property of $\epsilonTimesname$ 
  %and congruence laws for  $\transTimesSymb$
\label{r:cart:umpTrans}}
\end{mdframed}
\end{minipage}
}
\begin{manyfigcap}
Rules for $\langCart(\graph)$.
\end{manyfigcap}
\end{figure*}

\section{A type theory for fp-bicategories} \label{sec:fp-bicats-type-theory}

We extend the basic language $\langBicat$ to a type theory $\langCart$ with
finite products. The addition of products enables us to use arbitrary
contexts, defined as finite lists of variable-and-type pairs in which variable
names must not be repeated.  The underlying signature is therefore a
\mbox{2-multigraph} and the bicategorical structure of $\langBicat$ is
extended to a biclone.

The additional structure required for products is synthesised directly from
the cases in Definition~\ref{def:fp-bicat}.  These additional rules are
collected in Figures~\ref{r:cart:products-terms}--\ref{r:cart:umpTrans}. 
As for $\langBicat$, we work up to $\alpha$-equivalence of terms and rewrites.
The well-formedness properties of $\langBicat$ extend to $\langCart$. 

For every $n \in \Nat$ and types~$A_1, \dots, A_n$ we introduce a product type
$\prod_n(A_1, \dots, A_n)$;  
the case $n=0$ yields the unit type~$\unittype$. 
We fix a set of base types $S$ and let the set of types $\allProdTypes{S}$ 
be generated by the grammar 
\[\begin{array}{rcl}
A_1, \dots, A_n %\in \allProdTypes S
& ::= & X \in S 
\\[1mm]
& \st & \textstyle{\prod_n}(A_1, \dots, A_n)  \qquad (n \in \Nat)
\end{array}\]
On top of this, we fix a \mbox{2-multigraph}~$\graph$ with 
$\graph_0=\allProdTypes{S}$. 

We generally abuse notation by adopting the conventions of
Notation~\ref{not:products}(\ref{c:types}).  
For the biuniversal arrows we introduce distinguished constants~$\pi_k(p) :
A_k$~($k = 1, \dots, n$) for every context
\mbox{$\big(p : \prodop_n (A_1, \dots, A_n)\big)$}. As for
adjunction~(\ref{diag:fp-bicategory}),
we postulate an operator $\pair{-, \dots, =}$, a family of rewrites
\begin{equation*} \label{epsilonTimes}
%\!
  \epsilonTimesname_{t_1,\ldots,t_n}
  = %\!=\!
  \big(
    \epsilonTimesInd{k}{t_1, \dots, t_n} 
    : %\!:\!  
    \hcomp{\pi_k}{\pair{t_1, \dots, t_n}} 
    \To %\!\To\! 
    t_k
  \big)_{k = 1, \dots, n}
\end{equation*}
(recall Notation~\ref{not:BicatConventions}(\ref{c:constants})) for every
family of derivable terms $(\Gamma \vdash t_i : A_i)_{i = 1, \dots, n}$, and
provide a natural bijective correspondence between rewrites as follows:
\begin{equation}  \label{tree:cart:nesting}
	\begin{bprooftree} 
		\AxiomC{$\alpha_i: \hcomp{\pi_i}{u} \To t_i \quad (i = 1, \dots, n)$} 
		\doubleLine 
		\UnaryInfC{$\transTimes{\alpha_1,\ldots,\alpha_n}
      : u \To \pair{t_1, \dots, t_n}$} 
	\end{bprooftree} 
\end{equation}
The rules of Figure~\ref{r:cart:umpTrans} achieve this by making
$\epsilonTimesname$ universal: this is precisely the content of
Lemma~\ref{lem:cart:trans-ump} below.  These rules define \emph{lax} products.
To obtain the adjoint equivalence~(\ref{diag:fp-bicategory}) one introduces
explicit inverses for the unit and counit.  In this extended abstract these
are omitted.

\begin{myremark} \label{rem:cart:nesting}
The product structure arises from two \emph{nested} universal transposition
structures.  We conjecture that a calculus for
fp-\emph{tri}categories~(resp.~fp-$\infty$-categories) would have
\emph{three}~(resp.~a countably infinite tower of) such 
correspondences. 
\end{myremark} 

\setlength{\floatsep}{5pt plus 1.0pt minus 2.0pt}
 
\begin{figure*}
\begin{minipage}{\textwidth}

\centering
{\small
\begin{minipage}{\textwidth}
\begin{mdframed}
\begin{minipage}{\textwidth}
\centering 
	
			\unaryRule{(\Gamma \vdash \id_{t_i} : \rewrite{t_i}{t_i} : A_i)_{i = 1, \dots, n}}
					{\Gamma \vdash \pair{\id_{t_1}, \dots, \id_{t_n}} \equiv \id_{\pair{t_1, \dots, t_n}} : \rewrite{\pair{t_1, \dots, t_n}}{\pair{t_1, \dots, t_n}} : \prodop_n(A_1, \dots, A_n)}
					{}
{\footnotesize
			\binaryRule{(\Gamma \vdash \tau_i' : \rewrite{t_i'}{t_i''} : A_i)_{i = 1, \dots, n}}
					{(\Gamma \vdash \tau_i : \rewrite{t_i}{t_i'} : A_i)_{i = 1, \dots, n}}
					{\Gamma \vdash \pair{\tau_1', \dots, \tau_n'} \vert \pair{\tau_1, \dots, \tau_n} \equiv \pair{\tau_1' \vert \tau_1, \dots, \tau_n' \vert \tau_n} : \rewrite{\pair{t_1, \dots, t_n}}{\pair{t_1'', \dots, t_n''}}: \prodop_n(A_1, \dots, A_n)}
					{}}
\end{minipage}

\begin{minipage}{\textwidth}
\centering 

			\unaryRule	{\Gamma \vdash \sigma : \rewrite{u}{u'} : \prodop_n(A_1, \dots, A_n)}
						{\Gamma \vdash \etaTimes{u'} \vert \sigma \equiv \pair{\hcomp{\pi_1}{\sigma}, \dots, \hcomp{\pi_n}{\sigma}} \vert \etaTimes{u} : \rewrite{u}{\pair{\hcomp{\pi_1}{u'}, \dots, \hcomp{\pi_n}{u'}}} : \prodop_n(A_1, \dots, A_n)}
						{$\etaTimes{}$-nat}
		
		\unaryRule	{(\Gamma \vdash \tau_i : \rewrite{t_i}{t_i'} : A_i)_{i = 1, \dots, n}}
					{\Gamma \vdash \epsilonTimesInd{k}{t_1', \dots, t_n'} \vert \hcomp{\pi_k}{\pair{\tau_1, \dots, \tau_n}} \equiv \tau_k \vert \epsilonTimesInd{k}{t_1, \dots, t_n}  : \rewrite{\hcomp{\pi_k}{\pair{t_1, \dots, t_n}}}{t_k} : A_k }
					{$\epsilonTimesInd{k}{}$-nat $(1 \leq k \leq n)$}
\end{minipage}
	
\begin{minipage}{\textwidth}
\centering 
	{
	\unaryRule{\Gamma \vdash \pair{t_1, \dots, t_n} : \prodop_n(A_1, \dots, A_n)}
		{\Gamma \vdash \pair{\epsilonTimesInd{1}{\ind{t}}, \dots, \epsilonTimesInd{n}{\ind{t}}} \vert \etaTimes{\pair{\ind{t}}} \equiv \id_{\pair{\ind{t}}} : \rewrite{\pair{\ind{t}}}{\pair{\ind{t}}} : \prodop_n(A_1, \dots, A_n)}
		{triangle-law-1}
	}
	\unaryRule{\Gamma \vdash \hcomp{\pi_k}{u} : A_k}
				{\Gamma \vdash \epsilonTimesInd{k}{t_1, \dots, t_n} \vert \hcomp{\pi_k}{\etaTimes{u}} \equiv \id_{\hcomp{\pi_k}{u}} : \rewrite{\hcomp{\pi_k}{u}}{\hcomp{\pi_k}{u}} : A_k}
				{triangle-law-2 $(1 \leq k \leq n)$}
\end{minipage}

{
\caption{Admissible rules for $\langCart(\graph)$\label{fig:cart:admissible-rules}}
}
\end{mdframed}
\vspace{1.5em}
\end{minipage}
}
\end{minipage}
\end{figure*}

\subsection{The product structure of $\langCart$}

The product structure derives from the following universal property
of~$\epsilonTimesname$.

\begin{mylemma} \label{lem:cart:trans-ump} 
If the judgements 
$( \Gamma \vdash \alpha_i : \rewrite{\hcomp{\pi_i}{u}}{t_i} : A_i
   )_{i = 1, \dots, n}$ 
are derivable in 
$\langCart(\graph)$ then 
$\transTimes{\alpha_1, \dots, \alpha_n} : u \To \pair{t_1, \dots, t_n}$ is the
unique rewrite $\gamma$ (modulo \mbox{$\alpha{\equiv}$-equivalence}) such that
the following equality is derivable for $k = 1, \dots, n$: 
\[
\Gamma \vdash 
  \epsilonTimesInd{k}{t_1, \dots, t_n} \vert \hcomp{\pi_k}{\gamma} 
  \equiv \alpha_k 
  : \rewrite{\hcomp{\pi_i}{u}}{t_k} : A_k
\]
\end{mylemma}

From this lemma it follows that the $\pair{-, \dots,=}$ operator extends to a
functor on rewrites, and that one may define the unit of
adjunction~(\ref{diag:fp-bicategory}) by applying the universal property to
the identity. 

\begin{mydefn}  \label{def:fp:unit} \quad
\begin{enumerate}
\item 
  For every family of derivable rewrites 
  $(\Gamma \vdash \tau_i : \rewrite{t_i}{t_i'} : A_i)_{i=1,\dots, n}$ 
  define 
  $\pair{\tau_1, \dots, \tau_n} 
   : \pair{t_1, \dots, t_n} \To \pair{t_1', \dots, t_n'}$ 
  to be the rewrite 
  $\transTimes{
       \tau_1 
     \vert 
       \epsilonTimesInd{1}{t_1, \dots, t_n}, \dots, \tau_n 
     \vert 
       \epsilonTimesInd{n}{t_1, \dots, t_n}
   }$ 
  in context $\Gamma$. 

\item 
  For every derivable term $\Gamma \vdash t : \prodop_n(A_1, \dots, A_n)$ 
  define the unit 
  $\etaTimes{t} : t \To \pair{\hcomp{\pi_1}{t}, \dots, \hcomp{\pi_n}{t}}$ to
  be the rewrite 
  $\transTimes{\id_{\hcomp{\pi_1}{t}}, \dots, \id_{\hcomp{\pi_n}{t}}}$ in
  context $\Gamma$.  
\end{enumerate}
\end{mydefn}

Likewise, one obtains naturality and the triangle laws relating the
unit~$\etaTimes{}$ and 
counit~$\epsilonTimesname=(\epsilonTimesInd{1}{}, \dots,
\epsilonTimesInd{n}{})$, thereby recovering a presentation of products in the
style of~\cite{Seely1987, Hilken1996, Hirschowitz2013}. These admissible rules
are collected in Figure~\ref{fig:cart:admissible-rules}. 

\begin{figure*}[!h]
\begin{minipage}{\textwidth}
{
\begin{mdframed}
\begin{align*} 
h\sem{X} 
  &:= h(X) \qquad \text{for $X$ a base type}
\\
h\sem{\prodop_n(B_1, \dots, B_n)} 
  &:= \prodop_{i=1}^n h\sem{B_i} 
\\
h\sem{\exptype{A}{B}} 
  &:= \expobj{h\sem{A}}{h\sem{B}}  
\\
h\sem{x_1 : A_1, \dots, x_n : A_n} 
  &:= \prodop_{i=1}^n h\sem{A_i} 
      \qquad \text{on $n$-ary contexts} 
\\[5pt]
h\sem{x_1 : A_1, \dots, x_n : A_n \vdash x_k :A_i} 
  &:= \pi_k^{A_1, \dots, A_n} 
\\ 
h\sem{x_1 : A_1, \dots, x_n : A_n \vdash c(x_1, \dots, x_n) : B} 
  &:= h(c) 
      \qquad \text{ for $c \in \graph(\ind{A};B)$}
\\ 
h\sem{\Delta \vdash \hcomp{t}{x_i^{A_i} \mapsto u_i}_{i=1}^n : B} 
  &:= h\sem{x_1 : A_1, \dots, x_n : A_n \vdash t : B} 
\\&\hspace{20mm} \circ \tup{h\sem{\Delta \vdash u_i : A_i}}_{i = 1, \dots, n} 
\\
h\sem{\Gamma \vdash \pair{t_1, \dots, t_n} : \prodop_n(B_1, \dots, B_n)} 
&:= \tup{h\sem{\Gamma \vdash t_1 : B_1,},\dots, h\sem{\Gamma \vdash t_n : B_n}} \\
h\sem{p : \prodop_n(B_1, \dots, B_n) \vdash \pi_k(p) : B_k} 
&:= \pi_k^{B_1, \dots, B_n} 
\\
h\sem{f : \exptype{A}{B}, a : A \vdash \evalterm(f,a) : B} 
&:= \eval_{A,B} 
\\
h\sem{\Gamma \vdash \lam{x}{t} : \exptype{A}{B}} 
  &:= \lambda\big(h\sem{\Gamma, x : A \vdash t : B}\circ\simeq\big) 
\\[5pt]
h\sem{\Gamma \vdash \id_t : \rewrite{t}{t} : B} 
  &:= \id_{h\sem{\Gamma \vdash t : B}} 
\\
h\sem{x_1 : A_1, \dots, x_n : A_n 
      \vdash \sigma(\ind{x}) : \rewrite{c(\ind{x})}{c'(\ind{x})} : B} 
&:= h(\sigma) \qquad \text{ for $\sigma \in \graph(\ind{A}, B)(c,c')$}
\\
h\sem{\Gamma 
      \vdash \epsilonTimesInd{k}{t_1, \dots, t_n} 
           : \rewrite{\hcomp{\pi_k}{\pair{t_1, \dots, t_n}}}{t_k} : B_k} 
  &:= \epsilonTimesInd{k}{h\sem{t_1}, \dots, h\sem{t_n}} 
\\
h\sem{\Gamma 
      \vdash \transTimes{\alpha_1, \dots, \alpha_n} 
      : \rewrite{u}{\pair{t_1, \dots, t_n}} 
      : \prodop_n(B_1, \dots, B_n)} 
  &:= \transTimes
      {  h\sem{\Gamma \vdash \alpha_i : \rewrite{\hcomp{\pi_i}{u}}{t_i} : B_i}
         }_{i=1, \dots, n} 
\\
h\sem{\Gamma, x : A 
      \vdash \epsilonExpRewr{t} 
           : \rewrite{\genevalterm{\wkn{(\lam{x}{t})}{x}, x}}{t} : B} 
&:= \epsilon_{(h\sem{\Gamma, x : A \vdash t : B}\circ\simeq)} 
\\
h\sem{\Gamma 
      \vdash \transExp{x \bind \alpha} 
           : \rewrite{u}{\lam{x}{t}} 
           : \exptype{A}{B}} 
  &:= \transExp{
        h\sem{\Gamma, x : A 
              \vdash \alpha 
              : \rewrite{\genevalterm{\wkn{u}{x}, x}}{t} : B}
        \circ\simeq
    } 
\\
h\sem{\Gamma \vdash \tau' \vert \tau : \rewrite{t}{t''} : B} 
  &:= h\sem{\Gamma \vdash \tau' : \rewrite{t'}{t''} : B} 
      \vert 
      h\sem{\Gamma \vdash \tau : \rewrite{t}{t'} : B} 
\\
h\sem{ \Delta 
       \vdash 
       \hcomp{\tau}{x_i^{A_i} \mapsto \sigma_i}_{i= 1}^n 
       : \rewrite
           {\hcomp{t}{x_i^{A_i} \mapsto u_i}_{i= 1}^n}
           {\hcomp{t'}{x_i^{A_i} \mapsto u_i'}_{i= 1}^n} 
         : B} 
  &:= h\sem{x_1 : A_1, \dots, x_n : A_n \vdash \tau : \rewrite{t}{t'} : B} 
\\ 
	&\hspace{25mm} \circ 
  \tup{ h\sem{\Delta \vdash \sigma_i : \rewrite{u_i}{u_i'} : A_i}
        }_{i=1, \dots, n}
\end{align*} 
\caption{Bicategorical semantics \label{fig:extension-pseudofunctor}}
\end{mdframed}
}
\end{minipage}
\end{figure*}

\subsection{The syntactic model for $\langCart$} \label{sec:fp:syntactic-model}

We construct the syntactic model for $\langCart$ and prove the freeness
universal property establishing that it is the internal language of
\mbox{fp-bicategories}. We begin with a model in which we allow arbitrary
variables and consider all contexts. We then restrict to unary contexts and a
single named variable~(\cf~Constructions~\ref{constr:bicat-termcat} 
and~\ref{constr:bicat-termcat-fixedvar}) in order to obtain a strict universal
property~(\cf~Theorem~\ref{prop:BicatTermModelUMP}). 

\begin{myconstr} \label{constr:cart-termcat} 
Define a bicategory $\termCatTimes(\graph)$ as follows. The objects are
contexts $\Gamma, \Delta, \dots$. The 1-cells
${\Gamma \to (y_j : B_j)_{j = 1, \dots, m}}$ are $m$-tuples of
\mbox{$\alpha$-equivalence} classes of terms 
$(\Gamma \vdash t_j :B_j)_{j= 1, \dots, m}$ derivable in $\langCart(\graph)$;
the \mbox{2-cells}
are $m$-tuples of $\alpha{\equiv}$-equivalence classes of rewrites
${(\Gamma \vdash \tau : \rewrite{t_j}{t'_j} : B_j)_{j= 1, \dots, m}}$. 

Vertical composition is given pointwise by the $\vert$ operation, and
horizontal composition by explicit substitution: 
\begin{small}
\begin{align*} 
(t_1, \dots, t_l), (u_1, \dots, u_m) 
  \mapsto (\hcomp{t_1}{x_i \mapsto u_i}, \dots, \hcomp{t_l}{x_i \mapsto u_i}) 
\\[2mm]
(\tau_1, \dots, \tau_l), (\sigma_1, \dots, \sigma_m) 
  \mapsto ( \hcomp{\tau_1}{x_i \mapsto \sigma_i}, 
            \dots, 
            \hcomp{\tau_l}{x_i \mapsto \sigma_i})
\end{align*} 
\end{small}
The identity on $\Delta = (y_j : B_j)_{j = 1, \dots, m}$ is given by
the \rulename{var} rule \mbox{$(\Delta \vdash y_j : B_j)_{j = 1, \dots, m}$}.
The structural isomorphisms $\l, \r$ and $\a$ are given pointwise by
$\proj{}$, $\subid{}^{-1}$ and $\assoc{}$, respectively.  
\end{myconstr} 

The first step in showing that $\termCatTimes(\graph)$ is an
\mbox{fp-bicategory} is constructing $n$-ary products of unary contexts.  Much
of the work required is contained in the admissible rules of
Figure~\ref{fig:cart:admissible-rules}.  For example, there is an adjoint
equivalence 
${\termCatTimes(\graph)\big( (x : X), (p : \prodop_n(A_1, \dots, A_n)\big)}
 \simeq 
 {\prodop_{i=1}^n \termCatTimes(\graph)\big( (x : X), (x_i : A_i)\big)}$
defining products of unary contexts in $\termCatTimes(\graph)$ with unit
$\etaTimes{}$ and counit 
$\epsilonTimesname=(\epsilonTimesInd{1}{}, \dots, \epsilonTimesInd{n}{})$.

\begin{mylemma} \label{lem:fp-bicat:equivalences} 
For any 2-multigraph $\graph$, the following holds in $\termCatTimes(\graph)$.
\begin{enumerate}
\item 
  For any $n \in \Nat$ the $n$-ary product $\prod_{i=1}^n (x_i : A_i)$ exists
  and is given by $(p : \prod_{i=1}^n A_i)$.  
\item \label{c:context-to-unary}
  For any context $\Gamma = (x_1 : A_1, \dots, x_n : A_n)$ there exists an
  adjoint equivalence $\Gamma \leftrightarrows \big(p : \prodop_n(A_1, \dots,
  A_n)\big)$.  
\end{enumerate}
\end{mylemma}

\begin{myremark} 
The existence of the adjoint equivalence of
Lemma~\ref{lem:fp-bicat:equivalences}(\ref{c:context-to-unary}) in
$\termCatTimes(\graph)$ is equivalent to the existence of a
\Def{pseudo-universal} multimap
$\seq{A_1, \dots, A_n} \to \prodop_n(A_1, \dots, A_n)$ in the biclone
associated to $\langCart(\graph)$ (\cf~\cite{Hermida2000}).  
\end{myremark}

We define products of arbitrary contexts using
Remark~\ref{rem:nAryProductsDeterminedUpToEquiv} and the preceding lemma.
Define the product of 
\[
(x_i^{(1)} : A_i^{(1)})_{i = 1, \dots, m_1} 
\times \dots \times 
(x_i^{(n)} : A_i^{(n)})_{i = 1, \dots, m_n}
\]
to be the product 
$(p_1 : \prodop_{i=1}^{m_1} A_i^{(1)}) 
 \times \dots \times 
 (p_n : \prodop_{i=1}^{m_n} A_i^{(n)})$ 
of unary contexts.

\begin{mycor}  \label{cor:cart-termcat-fp}
For any 2-multigraph $\graph$, the syntactic model $\termCatTimes(\graph)$ of
$\langCart(\graph)$ is an fp-bicategory.  
\end{mycor} 

$\termCatTimes(\graph)$ contains redundancy in two ways: as well as the
equivalent objects given by bijectively renaming variables in contexts, 
every context $(x_i : A_i)_{i = 1,\dots, n}$ is equivalent to the context 
$(p : \prod_{i=1}^n A_i)$. To obtain a strict universal property we cut out
such multiplicities (\cf~Construction~\ref{constr:bicat-termcat-fixedvar}).  

\begin{myconstr}
Let $\termCatTimesSlim(\graph)$ denote the bicategory with objects unary
contexts $(x : A)$ for $x$ a \emph{fixed} variable and horizontal and vertical
composition operations as in $\termCatTimes(\graph)$.  
\end{myconstr}

The product structure of $\termCatTimes(\graph)$ restricts to
$\termCatTimesSlim(\graph)$ and the two bicategories are 
equivalent as fp-bicategories.
To obtain the freeness universal property for $\termCatTimes(\graph)$ with
respect to $2$-multigraphs $\graph$, it suffices to show that
$\termCatTimesSlim(\graph)$ is the free \mbox{fp-bicategory} for
$2$-graphs~$\graph$.  Universal extensions are induced by the semantics in
Figure~\ref{fig:extension-pseudofunctor} restricted to
$\termCatTimesSlim(\graph)$.

\begin{mythm} 
Let $\graph$ be a 2-graph with $\graph_0=\allProdTypes{S}$ for a set of base
types $S$.
For every \mbox{fp-bicategory} $\baseCat$
and \mbox{2-graph} homomorphism $h : \graph \to \baseCat$ such that
$h\big(\Pi_n(A_1,\ldots,A_n)\big)=\Pi_n(hA_1,\ldots,hA_n)$,
there exists a unique strict fp-pseudofunctor 
$h\semext : {\termCatTimesSlim(\graph) \to \baseCat}$ such that 
$h\semext \circ \inc = h$, where 
$\inc : \graph \hookrightarrow \termCatTimesSlim(\graph)$ denotes the
inclusion.
\end{mythm} 

\begin{figure*}[!h]
{\small
\begin{minipage}{\textwidth}
\begin{mdframed}
\centering
\unaryRule 	{\Gamma, x : A \vdash t : B} 
			{\Gamma \vdash \lam{x}{t} : \exptype{A}{B}} 
			{lam} 
\unaryRule 	{\phantom{a}} 
			{f : \exptype{A}{B}, a : A \vdash \evalterm(f,a) : B} 
			{eval} 
\vspace{-\treeskip}
\caption{Terms for cartesian closed structure \label{r:ccc:terms}}
\end{mdframed}
\end{minipage}

\begin{minipage}{\textwidth}
\begin{mdframed}
\centering
\begin{bprooftree} 
\AxiomC{$\Gamma, x : A \vdash t : B$} 
\RightLabel{\scriptsize $\epsilonExpRewr{}$-intro} 
\UnaryInfC{$\Gamma, x : A \vdash \epsilonExpRewr{t} : \rewrite{\genevalterm{\wkn{(\lam{x}{t})}{x}, x}}{t} : B$} 
\end{bprooftree}
\begin{bprooftree} 
\AxiomC{$\Gamma, x : A \vdash t : B$ \:\quad\: $\Gamma \vdash u : \exptype{A}{B}$} 
\noLine 
\UnaryInfC{$\Gamma, x : A \vdash \alpha : \rewrite{\genevalterm{\wkn{u}{x}, x}}{t} : B$} 
\RightLabel{{\scriptsize $\transExpSymb$-intro}} 
\UnaryInfC{$\Gamma \vdash \transExp{x \bind \alpha} : \rewrite{u}{\lam{x}{t}} : \exptype{A}{B}$} 
\end{bprooftree} \vspace{\treeskip} 
\caption{Rewrites for cartesian closed structure \label{r:ccc:rewrites}}
\end{mdframed}
\end{minipage}

\begin{minipage}{\textwidth}
\begin{mdframed}
\centering
\begin{bprooftree}
\AxiomC{$\Gamma, x : A \vdash \alpha : \rewrite{\genevalterm{\wkn{u}{x}, x}}{t} : B$}
\RightLabel{\scriptsize U1}
\UnaryInfC{$\Gamma, x : A \vdash \alpha \equiv \epsilonExpRewr{t} \vert \genevalterm{\wkn{\transExp{x \bind \alpha}}{x}, x} : \rewrite{\genevalterm{\wkn{u}{x}, x}}{t} : B$}
\end{bprooftree}\vspace{\treeskip}

\begin{bprooftree}
\AxiomC{$\Gamma \vdash \gamma : \rewrite{u}{\lam{x}{t}} : \exptype{A}{B}$}
\RightLabel{\scriptsize U2}
\UnaryInfC{$\Gamma \vdash \gamma \equiv \transExp{x \bind \epsilonExpRewr{t} \vert \genevalterm{\wkn{\gamma}{x}, x}} : \rewrite{u}{\lam{x}{t}} : \exptype{A}{B} $}
\end{bprooftree}\vspace{\treeskip}
%%%\input{rules/ccc/extra-congruence}
%\binaryRule	{\Gamma, y : B \vdash \alpha \equiv \alpha' : \rewrite{\genevalterm{\wkn{u}{y}, y}}{t} : C}
%			{\Gamma \vdash u : \exptype{B}{C}}			
%			{\Gamma \vdash \transExp{y \dot \alpha} \equiv \transExp{y \dot \alpha'} : \rewrite{u}{\lam{y}{t}} : \exptype{B}{C}}
%			{} 
			
\unaryRule	{\Gamma, x : A \vdash \alpha \equiv \alpha' : \rewrite{\genevalterm{\wkn{u}{x}, x}}{t} : B}		
			{\Gamma \vdash \transExp{x \bind \alpha} \equiv \transExp{x \bind \alpha'} : \rewrite{u}{\lam{x}{t}} : \exptype{A}{B}}
			{cong} 
\caption{Universal property of $\epsilonExpRewrName$
  \label{r:ccc:umpTrans}}
\end{mdframed}
\end{minipage}
}

\begin{manyfigcap}
Rules for $\langCartClosed(\graph)$.
\end{manyfigcap}

\end{figure*}

\section{Cartesian closed bicategories} \label{sec:cc-bicats}

To give a cartesian closed structure on an fp-bicategory $\baseCat$ is to
specify a biadjunction $(-) \times A \dashv (\expobj{A}{-})$ for each
object~$A \in \baseCat$. Cartesian closed bicategories are the objects of a
tricategory $\CartClosedBicat$. 

\begin{mydefn} \label{def:cc-bicat}
A \Def{cartesian closed bicategory} or \mbox{\Def{CC-bicategory}} is an
fp-bicategory $\big(\baseCat, \Pi_n(-)\big)$ equipped with the following data
for every $A, B \in \baseCat$: 
\begin{enumerate}
\item \label{c:cc-bicat-obj} 
an \Def{exponential} object $(\expobj{A}{B})$, 

\item \label{c:cc-bicat-map} 
an \Def{evaluation} \mbox{1-cell} 
$\eval_{A,B} : (\expobj{A}{B}) \times A \to B$, 

\item \label{c:cc-bicat-adj-equiv} 
for every $X \in \baseCat$ an adjoint equivalence 
\begin{equation} \label{diag:cc-bicat}
\begin{tikzcd} 
\baseCat(X, \expobj{A}{B}) 
\arrow[bend left = 20]{r}{\eval_{A,B} \circ (- \times A)} 
\arrow[phantom]{r}[xshift=0em]{\adjUp} & 
\baseCat(X \times A, B) \arrow[bend left = 20]{l}{\lambda} 
\end{tikzcd} 
\end{equation}
\end{enumerate}
\end{mydefn}

\begin{myremark} \label{rem:exponentials-up-to-equiv}
The uniqueness of exponentials up to equivalence manifests itself in the same
way as for products.  For instance, given an adjoint 
equivalence~$e: E \simeq (\expobj{A}{B}): f$, the object $E$ inherits an
exponential structure by composition with $e$ and $f$
(\cf~Remark~\ref{rem:nAryProductsDeterminedUpToEquiv}).
\end{myremark}

As for products, we choose to define the adjunction~(\ref{diag:cc-bicat}) by
characterising its counit~$\epsilonExp$ as a universal arrow (\cf~the
discussion after Notation~\ref{not:products}).
Thus, for every 1-cell ${t : X \times A \to B}$ we require a \mbox{1-cell}
${\lambda t : X \to (\expobj{A}{B})}$ and a \mbox{2-cell}
$\epsilon_t : \eval_{A,B} \circ (\lambda t \times A) \To t$ 
that is universal in the sense that for any \mbox{2-cell} 
$\alpha : \eval_{A,B} \circ (u \times A) \To t$ there exists a unique 
\mbox{2-cell} $\transExp{\alpha} : u \To \lambda t$ such that 
$\epsilon_t \vert \big(\eval_{A,B} \circ (\transExp{\alpha} \times A)\big) 
 = \alpha$. 

\begin{mydefn} 
A \Def{cartesian closed pseudofunctor} or \mbox{\Def{CC-pseudofunctor}}
between CC-bicategories $\big(\baseCat, \Pi_n(-), {\altTo}\big)$ and
\mbox{$\big(\altCat, \Pi_n(-), {\altTo}\big)$} is an \mbox{fp-pseudofunctor}
$(F, \prodPres)$ equipped with equivalences
\[\big(
  F(\expobj{A}{B}) ,
  (\expobj{FA}{FB}) , 
  \evBar_{A,B} ,
  \expPres_{A,B}
\big)\]
for every $A, B \in \baseCat$, where 
$\evBar_{A,B} :  F(\expobj{A}{B}) \to (\expobj{FA}{FB})$ is the transpose of
\[
F(\eval_{A,B}) \circ \prodPres_{\scriptsizeexpobj{A}{B}, A} 
 : F(\expobj{A}{B}) \times FA \to FB
\]
 
A CC-pseudofunctor $(F, \prodPres, \expPres)$ is \Def{strict} if 
$(F, \prodPres)$ is strict and satisfies
\begin{align*}
F(\expobj{A}{B}) &= (\expobj{FA}{FB}) 
\\
F(\eval_{FA,FB}) &= \eval_{FA,FB} 
\\
F(\lambda t) &= \lambda(Ft) 
\\
F(\epsilonExp_t) &=\epsilonExp_{Ft} 
\\
\expPres_{A,B} &= \Id_{\scriptsizeexpobj{FA}{FB}}
\end{align*}
and the adjoint equivalences are canonically induced by the %canonical
\mbox{$2$-cells}
$\transExp{\kappa}
 : \Id_{\scriptsizeexpobj{FA}{FB}}
   \XRA\iso
   \lambda(\eval_{FA,FB})$
where $\kappa$ is the following composite of canonical $2$-cells 
\[
\eval_{FA,FB}\circ(\Id\times FA)\iso\eval_{FA,FB}\circ\Id\iso\eval_{FA,FB}
\]
\end{mydefn}

\begin{figure*}[!h]
\begin{minipage}{\textwidth}
{\small
\begin{mdframed}
\begin{minipage}{\textwidth}
%\begin{mdframed}
\centering 

\unaryRule {\Gamma, x : A \vdash t : B} 
{\Gamma \vdash \lam{x}{\id_t} \equiv \id_{\lam{x}{t}} : \rewrite{\lam{x}{t}}{\lam{x}{t}} : \exptype{A}{B}} 
{} 
\binaryRule {\Gamma, x : A \vdash \tau' : \rewrite{t'}{t''} : B} 
{\Gamma, x : A \vdash \tau : \rewrite{t}{t'} : B} 
{\Gamma \vdash \lam{x}{(\tau' \vert \tau)} \equiv (\lam{x}{\tau'}) \vert (\lam{x}{\tau}) : \rewrite{\lam{x}{t}}{\lam{x}{t''}} : \exptype{A}{B}} 
{} 
\end{minipage}

\begin{minipage}{\textwidth}
\vspace{-\treeskip}	
\centering
		\unaryRule {\Gamma \vdash \sigma : \rewrite{u}{u'} : \exptype{A}{B}} 
		{\Gamma \vdash \etaExp{u'} \vert \sigma \equiv \lam{x}{\genevalterm{\wkn{\sigma}{x}, x}} \vert \etaExp{u} : \rewrite{u}{\lam{x}{\hcomp{\evalterm}{\wkn{u'}{x}, x}}} : \exptype{A}{B}} 
		{$\etaExp{}$-nat} 
		\unaryRule {\Gamma, x : A \vdash \tau : \rewrite{t}{t'} : B} 
		{\Gamma, x : A \vdash \tau \vert \epsilonExpRewr{t} \equiv \epsilonExpRewr{t'} \vert \genevalterm{\wkn{(\lam{x}{\tau})}{x}, x} : \rewrite{ \genevalterm{\wkn{(\lam{x}{t})}{x}, x}}{t'} : B} 
		{$\epsilonExpRewr{}$-nat} 
\end{minipage}

\begin{minipage}{\textwidth}
\centering
\begin{mbox}
\centering 
	\unaryRule {\Gamma, x : A \vdash t : B} 
	{\Gamma \vdash (\lam{x}{\epsilonExpRewr{t}}) \vert \etaExp{t} \equiv \id_{\lam{x}{t}} : \rewrite{\lam{x}{t}}{\lam{x}{t}} : \exptype{A}{B}} 
	{triangle-law-1} 

	{\small
		\begin{bprooftree} 
			\AxiomC{$\Gamma \vdash u : \exptype{A}{B}$} 
			\RightLabel{{\scriptsize triangle-law-2}}
			\UnaryInfC{$\Gamma, x : A \vdash \epsilonExpRewr{\hcomp{\evalterm}{\wkn{u}{x}, x}} \vert \hcomp{\evalterm}{\wkn{\etaExp{u}}{x}, x} \equiv \id_{\hcomp{\evalterm}{\wkn{u}{x}, x}} : \rewrite{\hcomp{\evalterm}{\wkn{u}{x}, x}}{\hcomp{\evalterm}{\wkn{u}{x}, x}} : B$} 
		\end{bprooftree}\vspace{\treeskip} 
	} 
\end{mbox}
\end{minipage}

{
\caption{Admissible rules for $\langCartClosed(\graph)$\label{fig:ccc:admissible-rules}}
}
\end{mdframed}
}
\end{minipage}
\end{figure*}

\section{A type theory for cartesian closed bicategories} 
\label{sec:cc-bicats-type theory}

We extend $\langCart$ to the full language $\langCartClosed$ by synthesising
the additional rules from the cases of Definition~\ref{def:cc-bicat}; these
rules are Figures~\ref{r:ccc:terms}--\ref{r:ccc:umpTrans}. 

We extend the grammar for types with an arrow type former
$\exptype{(-)}{(=)}$.
We henceforth fix a set of base types~$S$, let $\allTypes{S}$ denote the set
of all types over $S$, and consider \mbox{2-multigraphs}~$\graph$ with set of
nodes $\nodes{\graph}=\allTypes{S}$. 

We postulate a constant $\evalterm(f,x)$ for every context 
\mbox{$(f : \exptype{A}{B}, x : A)$}; the usual application operation becomes
a derived rule:
\begin{equation*} \label{fig:application}
\begin{bprooftree}
	\AxiomC{$\Gamma \vdash t : \exptype{A}{B}$}
	\AxiomC{$\Gamma \vdash u : A$}
	\BinaryInfC{$\Gamma \vdash \hcomp{\evalterm}{t,u} : B$}
\end{bprooftree}
\end{equation*}
The adjunction~(\ref{diag:cc-bicat}) requires the functor~${(-) \times A}$ on
the syntactic model.  
As for STLC, this corresponds to context extension and an associated notion of
weakening.  The explicit nature of substitution in our type theory gives rise
to correspondingly explicit structural operations on contexts.

\begin{mydefn} 
A \Def{context renaming} 
\[
r : (x_i : A_i)_{i = 1,\dots, n} \to (y_j : B_j)_{j=1, \dots, m}
\]
is a map $r : \{x_1, \dots, x_n\} \to \{y_1, \dots, y_m\}$ such that 
$A_i = B_j$ whenever $r(x_i) = y_j$.
\end{mydefn}

It is immediate that for every context renaming $r : \Gamma \to \Delta$ the
following rules are admissible in
$\langCartClosed(\graph)$:
	\begin{center} 
		\begin{small}
		\begin{prooftree}
			\AxiomC{$\Gamma \vdash t : A$}
			\AxiomC{$r : \Gamma \to \Delta$}
			\BinaryInfC{$\Delta \vdash \hcomp{t}{x_i \mapsto r(x_i)} : A$}
		\end{prooftree}
		
		\begin{prooftree}
			\AxiomC{$\Gamma \vdash \tau: \rewrite{t}{t'} : A$}
			\AxiomC{$r : \Gamma \to \Delta$}
			\BinaryInfC{$\Delta \vdash \hcomp{\tau}{x_i \mapsto r(x_i)} 
        : \rewrite{\hcomp{t}{x_i \mapsto r(x_i)}}
                  {\hcomp{t'}{x_i \mapsto r(x_i)}} 
            : A$}
		\end{prooftree}
		\end{small}
	\end{center}

\begin{mynotation} \label{not:context-morphism}
For a context renaming $r$ we write $\hcomp{t}{r}$ and $\hcomp{\tau}{r}$ for
the terms and rewrites formed using the preceding admissible rules.
\end{mynotation}

Weakening arises by taking the inclusion of contexts 
${\mathrm{inc}_{x} : \Gamma \hookrightarrow (\Gamma, x : A)}$. 
For the counit of~(\ref{diag:cc-bicat}) we therefore postulate a rewrite
\[
\epsilonExpRewr{t} : \hcomp{\evalterm}{\wkn{(\lam{x}{t})}{x}, x} \To t 
\]
for every term $t$ typeable in a context extended by the variable~$x$. 

The encoding of the universal property of $\epsilonExpRewr t$ in the type
theory yields a bijective correspondence of rewrites modulo
$\alpha{\equiv}$-equivalence as follows (\cf~(\ref{tree:cart:nesting})):
\begin{equation*}  
\begin{bprooftree} 
\AxiomC{$(x : A) \quad \alpha : \hcomp{\evalterm}{\wkn{u}{x}, x} \To t : B$} 
\doubleLine 
\UnaryInfC{$\transExp{x \bind \alpha} : u \To \lam{x}{t} : \exptype{A}{B}$} 
\end{bprooftree} 
\end{equation*}
where we write $(x : A)$ to indicate that the variable $x$ of type $A$ is in
the context~(\cf~\cite{MartinLof}).
Our approach to achieving this matches that for products.  For every rewrite 
${\alpha : \hcomp{\evalterm}{\wkn{u}{x}, x} \To t}$ we introduce the rewrite
${\transExp{x \bind \alpha} : u \To \lam{x}{t}}$ and make it unique in
inducing a factorisation of $\alpha$ through $\epsilonExpRewr t$; 
the variable $x$ is bound by this operation. The rules governing this
construction are given in Figures~\ref{r:ccc:rewrites}--\ref{r:ccc:umpTrans};
these provide \Def{lax} exponentials (\cf~\cite{Hilken1996}).  Finally, one
requires explicit inverses for the unit and counit.  In this extended abstract
they are omitted.

\begin{myremark}
As for products (\cf~Remark~\ref{rem:cart:nesting}), we obtain a nesting of
two universal transposition structures.  In the same vein, we conjecture that
a calculus for cartesian closed \emph{tri}categories (cartesian closed
$\infty$-categories) would have \emph{three} (a countably infinite tower of)
of such correspondences.  
Indeed, this should extend to general type structures arising from weak
adjunctions.
\end{myremark} 

We continue to work up to $\alpha$-equivalence of terms and rewrites. 
The well-formedness properties of $\langBicat$ and $\langCart$ lift to
$\langCartClosed$. 

\subsection{Cartesian closed structure of $\langCartClosed$}

The $\epsilonExpRewr{}$-introduction rule (Figure~\ref{r:ccc:rewrites}) only
`evaluates' lambda abstractions at variables.  The general form of explicit
\mbox{$\beta$-reduction} is derivable.

\begin{mydefn}
For derivable terms \mbox{$\Gamma, x : A \vdash t : B$} and 
\mbox{$\Gamma \vdash u : A$} define 
\mbox{$\genEpsilonExp{x \bind t, u} 
  : \rewrite
      {\hcomp{\evalterm}{\lam{x}{t}, u}}
      {\hcomp{t}{\id_{\Gamma}, x \mapsto u}}$} 
to be the rewrite 
$\hcomp{\epsilonExpRewr{t}}{\id_{\Gamma}, x \mapsto u} \vert \tau$ in context
$\Gamma$ (recall Notation~\ref{not:context-morphism}), where $\tau$ is a
composite of structural isomorphisms.  
\end{mydefn}

As expected, the remaining exponential structure follows from the universal
property of $\epsilonExpRewr{}$ (\cf~Lemma~\ref{lem:cart:trans-ump}).

\begin{mylemma} \label{lem:ccc:trans-ump}
For every rewrite 
$({\Gamma, x : A} \vdash \alpha 
     : \rewrite{\hcomp{\evalterm}{\wkn{u}{x}, x}}{t}
         : B)$ 
in $\langCartClosed(\graph)$, the rewrite $\transExp{x \bind \alpha}$ is the
unique $\gamma$ (modulo $\alpha{\equiv}$-equivalence) such that 
\begin{small}
\begin{center}
$\Gamma, x : A \vdash \alpha \equiv \epsilonExpRewr{t} \vert \genevalterm{\wkn{\gamma}{x}, x} : \rewrite{\hcomp{\evalterm}{\wkn{u}{x}, x}}{t} : B$
\end{center}
\end{small}
\end{mylemma}

It follows that the lambda-abstraction operator extends to a functor, and we
obtain a  unit for the adjunction~(\ref{diag:cc-bicat}).

\begin{mydefn} \quad
\begin{enumerate}
\item 
For any derivable rewrite 
\mbox{$(\Gamma, x : A \vdash \tau : \rewrite{t}{t'} :B)$} define
$\lam{x}{\tau} : \rewrite{\lam{x}{t}}{\lam{x}{t'}}$ to be the rewrite
$\transExp{ x \bind \tau \vert \epsilonExpRewr{t}}$ in context $\Gamma$.

\item 
For any derivable term $(\Gamma \vdash u : \exptype{A}{B})$ define the unit
$\etaExp{u} : \rewrite{u}{\lam{x}{\genevalterm{\wkn{u}{x}, x}}}$ to be the
rewrite $\transExp{x \bind \id_{\hcomp{\evalterm}{\wkn{u}{x}, x}}}$ in context
$\Gamma$.
\end{enumerate}
\end{mydefn}

The unit $\etaExp{}$ and counit $\epsilonExpRewr{}$ satisfy naturality and
triangle laws,  collected in Figure~\ref{fig:ccc:admissible-rules}.  Thus we
recover the unit-counit presentation of both products and exponentials
\mbox{(\cf~\cite{Seely1987, Hilken1996, Hirschowitz2013})}. 

\subsection{The syntactic model for $\langCartClosed$} 

We finally turn to constructing the syntactic model for
$\langCartClosed(\graph)$ and proving its freeness universal property. The
construction of the syntactic model follows the pattern established by
$\langBicat$ and $\langCart$.

\begin{myconstr} \label{constr:ccc-termcat}  
Define a bicategory $\termCatCCC(\graph)$ as
follows. The objects are contexts $\Gamma, \Delta, \dots$. The 1-cells 
${\Gamma \to (y_j : B_j)_{j = 1, \dots, m}}$ are $m$-tuples of 
$\alpha$-equivalence classes of terms 
$(\Gamma \vdash t_j :B_j)_{j= 1, \dots, m}$ derivable in
$\langCartClosed(\graph)$; \mbox{2-cells} 
are $m$-tuples of $\alpha{\equiv}$-equivalence classes of rewrites
\mbox{$(\Gamma \vdash \tau : \rewrite{t_j}{t'_j} : B_j)_{j= 1, \dots, m}$}.
Horizontal and vertical composition are as
in~\mbox{Construction~\ref{constr:cart-termcat}}.  
\end{myconstr} 

$\termCatCCC(\graph)$ inherits the product structure of
$\termCatTimes(\graph)$.  To construct cartesian closed structure it suffices
to construct exponentials of unary contexts. Much of the work required is
contained in the admissible rules of Figure~\ref{fig:ccc:admissible-rules}.

\begin{mylemma} 
For unary contexts ${(x : A)}$, ${(y : B)}$, the exponential 
$\expobj{{(x : A)}}{{(y : B)}}$ in $\termCatCCC(\graph)$ exists and is given
by $(f : \exptype{A}{B})$.  
\end{mylemma}

Using Remark~\ref{rem:exponentials-up-to-equiv} and the preceding lemma, we
define the exponential object $(\expobj{\Gamma}{\Delta})$ for 
\mbox{$\Gamma := (x_i : A_i)_{i = 1, \dots, n}$} and 
\mbox{$\Delta := (y_j : B_j)_{j = 1, \dots, m}$} using the equivalent unary
contexts, as 
\begin{center}
$\expobj
   {\big(p : \prodop_n(A_1, \dots, A_n)\big)}
   {\big(q : \prodop_m(B_1, \dots, B_m)\big)}$.
\end{center} 

\begin{mycor} 
\qquad %%% HACK !!!
The syntactic model $\termCatCCC(\graph)$ is a \mbox{CC-bicategory}.  
\end{mycor}

From Sections~\ref{sec:bicat:syntactic-model} and~\ref{sec:fp:syntactic-model}
one might expect that restricting to a sub-bicategory with unary contexts and
a fixed variable name is sufficient to obtain the required freeness universal
property. This suggests the following definition.

\begin{myconstr}
Let $\termCatCCCSlim(\graph)$ denote the bicategory with objects unary
contexts $(x : A)$ for $x$ a \emph{fixed} variable and horizontal and vertical
composition operations as in $\termCatCCC(\graph)$.  
\end{myconstr}

The cartesian closed structure of $\termCatCCC(\graph)$ restricts to
$\termCatCCCSlim(\graph)$ and the two are equivalent as cartesian closed
bicategories.  However, one does not obtain a strict freeness universal
property.  
We recover uniqueness by restricting to cartesian closed bicategories in which
the product structure is strict.

\begin{mythm} 
Let $\graph$ be a 2-graph with $\nodes\graph=\allTypes S$ for a set of base
types $S$.  For every cartesian 2-category (\ie~\mbox{2-category} with
2-categorical products) with pseudo-exponentials $\altCat$ and every 2-graph
homomorphism $h : \graph \to \altCat$ such that
$h\big(\Pi_n(A_1,\ldots,A_n)\big)=\Pi_n(hA_1,\ldots,hA_n)$ and 
$h(\expobj A B) = (\expobj{hA}{hB})$, there exists a unique strict
\mbox{CC-pseudofunctor} $h\semext : \termCatCCCSlim(\graph) \to \altCat$ such
that $h\semext \circ \inc = h$, for 
$\inc : \graph \hookrightarrow \termCatCCCSlim(\graph)$ the inclusion.  
\end{mythm} 

Thus, one obtains the same result as for
Sections~\ref{sec:bicat:syntactic-model} and~\ref{sec:fp:syntactic-model},
albeit with a restricted freeness universal property.
This, modulo the coherence result of Power~\cite{Power1989bilimit} applied to
fp-bicategories, yields that 
\begin{center}
  $\langCartClosed$ is the internal language of\\[.5mm]
  cartesian closed bicategories. 
\end{center}

In fact, it is possible to adjust the definition of exponentials to obtain
another type theory for which the induced syntactic models are equivalent as
CC-bicategories to those above and satisfy a strict freeness universal
property with respect to arbitrary CC-bicategories.  This type theory is
however no longer in the spirit of STLC, and so will be presented elsewhere.  

\section*{Acknowledgements}

We thank Dylan McDermott and Ian Orton for helpful discussions on type
theories with weak substitution.

%%%\bibliographystyle{IEEEtran} 
%%%\bibliography{bicats_db} 

\begin{thebibliography}{10}
\providecommand{\url}[1]{#1}
\csname url@samestyle\endcsname
\providecommand{\newblock}{\relax}
\providecommand{\bibinfo}[2]{#2}
\providecommand{\BIBentrySTDinterwordspacing}{\spaceskip=0pt\relax}
\providecommand{\BIBentryALTinterwordstretchfactor}{4}
\providecommand{\BIBentryALTinterwordspacing}{\spaceskip=\fontdimen2\font plus
\BIBentryALTinterwordstretchfactor\fontdimen3\font minus
  \fontdimen4\font\relax}
\providecommand{\BIBforeignlanguage}[2]{{%
\expandafter\ifx\csname l@#1\endcsname\relax
\typeout{** WARNING: IEEEtran.bst: No hyphenation pattern has been}%
\typeout{** loaded for the language `#1'. Using the pattern for}%
\typeout{** the default language instead.}%
\else
\language=\csname l@#1\endcsname
\fi
#2}}
\providecommand{\BIBdecl}{\relax}
\BIBdecl

\bibitem{Benabou1967}
J.~B{\'e}nabou, ``Introduction to bicategories,'' in \emph{Reports of the
  Midwest Category Seminar}.\hskip 1em plus 0.5em minus 0.4em\relax Springer
  Berlin Heidelberg, 1967, pp. 1--77.

\bibitem{Street1995}
R.~Street, ``Categorical structures,'' in \emph{Handbook of Algebra},
  M.~Hazewinkel, Ed.\hskip 1em plus 0.5em minus 0.4em\relax Elsevier, 1995,
  vol.~1, ch.~15, pp. 529--577.

\bibitem{CattaniFioreWinskel}
{G.L. Cattani}, M.~Fiore, and G.~Winskel, ``A theory of recursive domains with
  applications to concurrency,'' in \emph{Proceedings of the 13th Annual IEEE
  Symposium on Logic in Computer Science}.\hskip 1em plus 0.5em minus
  0.4em\relax IEEE Computer Society, 1998, pp. 214--225.

\bibitem{CCRW2017}
\BIBentryALTinterwordspacing
S.~Castellan, P.~Clairambault, S.~Rideau, and G.~Winskel, ``Games and
  strategies as event structures,'' \emph{Logical Methods in Computer Science},
  vol.~13, no.~3, 2017. [Online]. Available:
  \url{https://doi.org/10.23638/LMCS-13(3:35)2017}
\BIBentrySTDinterwordspacing

\bibitem{Abbott2003}
M.~G. Abbott, ``Categories of containers,'' Ph.D. dissertation, University of
  Leicester, 2003.

\bibitem{Dagand2013}
P.-E. Dagand and C.~McBride, ``A categorical treatment of ornaments,'' in
  \emph{Proceedings of the 28th Annual ACM/IEEE Symposium on Logic in Computer
  Science}.\hskip 1em plus 0.5em minus 0.4em\relax IEEE Computer Society, 2013,
  pp. 530--539.

\bibitem{FioreSpecies}
M.~Fiore, N.~Gambino, M.~Hyland, and G.~Winskel, ``The cartesian closed
  bicategory of generalised species of structures,'' \emph{Journal of the
  London Mathematical Society}, vol.~77, no.~1, pp. 203--220, 2007.

\bibitem{Gambino2013}
N.~Gambino and J.~Kock, ``Polynomial functors and polynomial monads,''
  \emph{Mathematical Proceedings of the Cambridge Philosophical Society}, vol.
  154, no.~1, pp. 153--192, 2013.

\bibitem{Fiore2015}
\BIBentryALTinterwordspacing
M.~Fiore and A.~Joyal, ``Theory of para-toposes,'' talk at the \emph{Category
  Theory 2015 Conf.}, Departamento de Matematica, Universidade de Aveiro,
  Aveiro (Portugal), 2015. [Online]. Available:
  \url{http://sweet.ua.pt/dirk/ct2015/abstracts/fiore_m.pdf}
\BIBentrySTDinterwordspacing

\bibitem{Gambino2017}
N.~Gambino and A.~Joyal, ``On operads, bimodules and analytic functors,''
  \emph{Memoirs of the American Mathematical Society}, vol. 249, no. 1184, pp.
  153--192, 2017.

\bibitem{FGHW2017}
M.~Fiore, N.~Gambino, M.~Hyland, and G.~Winskel, ``Relative pseudomonads,
  {K}leisli bicategories, and substitution monoidal structures,'' \emph{Selecta
  Mathematica New Series}, 2017.

\bibitem{Gray1974}
J.~W. Gray, \emph{Formal Category Theory: Adjointness for 2-Categories}, ser.
  Lecture Notes in Mathematics.\hskip 1em plus 0.5em minus 0.4em\relax
  Springer, 1974, vol. 391.

\bibitem{MacLane1985}
S.~Mac~{L}ane and R.~Par{\'e}, ``Coherence for bicategories and indexed
  categories,'' \emph{Journal of Pure and Applied Algebra}, vol.~37, pp.
  59--80, 1985.

\bibitem{Power1989bilimit}
A.~J. Power, ``Coherence for bicategories with finite bilimits {I},'' in
  \emph{Categories in Computer Science and Logic: Proceedings of the
  AMS-IMS-SIAM Joint Summer Research Conference Held June 14-20, 1987}, J.~W.
  Gray and A.~Scedrov, Eds.\hskip 1em plus 0.5em minus 0.4em\relax American
  Mathematical Society, 1989, vol.~92, pp. 341--349.

\bibitem{elephant}
P.~T. Johnstone, \emph{Sketches of an Elephant}.\hskip 1em plus 0.5em minus
  0.4em\relax Oxford University Press, 2002.

\bibitem{Gambino2004}
N.~Gambino and M.~Hyland, ``Wellfounded trees and dependent polynomial
  functors,'' in \emph{Types for Proofs and Programs}, S.~Berardi, M.~Coppo,
  and F.~Damiani, Eds.\hskip 1em plus 0.5em minus 0.4em\relax Springer Berlin
  Heidelberg, 2004, pp. 210--225.

\bibitem{Lambek1985}
J.~Lambek, ``Cartesian closed categories and typed lambda- calculi,'' in
  \emph{Proceedings of the Thirteenth Spring School of the LITP on Combinators
  and Functional Programming Languages}.\hskip 1em plus 0.5em minus 0.4em\relax
  Springer-Verlag, 1986, pp. 136--175.

\bibitem{Makkai1977}
M.~Makkai and G.~Reyes, \emph{First Order Categorical Logic: Model-Theoretical
  Methods in the Theory of Topoi and Related Categories}.\hskip 1em plus 0.5em
  minus 0.4em\relax Springer, 1977.

\bibitem{Church1940}
A.~Church, ``A formulation of the simple theory of types,'' \emph{The Journal
  of Symbolic Logic}, vol.~5, no.~2, pp. 56--68, 1940.

\bibitem{FioreOpetopicBonn}
\BIBentryALTinterwordspacing
M.~Fiore, ``An algebraic combinatorial approach to opetopic structure,'' talk
  at the \emph{Seminar on Higher Structures, Program on Higher Structures in
  Geometry and Physics}, Max Planck Institute for Mathematics, Bonn (Germany),
  2016. [Online]. Available: \url{https://www.mpim-bonn.mpg.de/node/6586}
\BIBentrySTDinterwordspacing

\bibitem{Yamada2018}
\BIBentryALTinterwordspacing
N.~Yamada and S.~Abramsky, ``Dynamic game semantics,'' \emph{arXiv}, 2018.
  [Online]. Available: \url{https://arxiv.org/abs/1601.04147}
\BIBentrySTDinterwordspacing

\bibitem{Rydeheard1987}
D.~E. Rydeheard and J.~G. Stell, ``Foundations of equational deduction: A
  categorical treatment of equational proofs and unification algorithms,'' in
  \emph{Category Theory and Computer Science}, D.~H. Pitt, A.~Poign{\'e}, and
  D.~E. Rydeheard, Eds.\hskip 1em plus 0.5em minus 0.4em\relax Springer Berlin
  Heidelberg, 1987, pp. 114--139.

\bibitem{Power1989}
A.~J. Power, ``An abstract formulation for rewrite systems,'' in \emph{Category
  Theory and Computer Science}, D.~H. Pitt, D.~E. Rydeheard, P.~Dybjer, A.~M.
  Pitts, and A.~Poign{\'e}, Eds.\hskip 1em plus 0.5em minus 0.4em\relax
  Springer Berlin Heidelberg, 1989, pp. 300--312.

\bibitem{Seely1987}
R.~A.~G. Seely, ``Modelling computations: A 2-categorical framework,'' in
  \emph{Proceedings of the Second Annual IEEE Symp. on Logic in Computer
  Science}, D.~Gries, Ed.\hskip 1em plus 0.5em minus 0.4em\relax IEEE Computer
  Society Press, June 1987, pp. 65--71.

\bibitem{Hilken1996}
B.~P. Hilken, ``Towards a proof theory of rewriting: the simply typed
  2{$\lambda$}-calculus,'' \emph{Theoretical Computer Science}, vol. 170,
  no.~1, pp. 407--444, 1996.

\bibitem{Jay1995}
C.~B. Jay and N.~Ghani, ``The virtues of eta-expansion,'' \emph{Journal of
  Functional Programming}, vol.~5, no.~2, pp. 135--154, 1995.

\bibitem{Ghani1995}
N.~Ghani, ``Adjoint rewriting,'' Ph.D. dissertation, University of Edinburgh,
  1995.

\bibitem{Tabareau2011}
N.~Tabareau, ``Aspect oriented programming: A language for 2-categories,'' in
  \emph{Proceedings of the 10th International Workshop on Foundations of
  Aspect-oriented Languages}, ser. FOAL '11.\hskip 1em plus 0.5em minus
  0.4em\relax ACM, 2011, pp. 13--17.

\bibitem{Hirschowitz2013}
T.~Hirschowitz, ``Cartesian closed 2-categories and permutation equivalence in
  higher-order rewriting,'' \emph{Logical Methods in Computer Science}, vol.~9,
  pp. 1--22, 07 2013.

\bibitem{MartinLof}
P.~Martin-L\"{o}f, \emph{Intuitionistic Type Theory}.\hskip 1em plus 0.5em
  minus 0.4em\relax Bibliopolis, 1984.

\bibitem{Licata2011}
D.~R. Licata and R.~Harper, ``2-dimensional directed type theory,''
  \emph{Electronic Notes in Theoretical Computer Science}, vol. 276, pp.
  263--289, 2011, twenty-seventh Conference on the Mathematical Foundations of
  Programming Semantics (MFPS XXVII).

\bibitem{Curien1993}
P.-L. Curien, ``Substitution up to isomorphism,'' \emph{Fundam. Inf.}, vol.~19,
  no. 1-2, pp. 51--85, Sep. 1993.

\bibitem{hottbook}
{The Univalent Foundations Program}, \emph{Homotopy Type Theory: Univalent
  Foundations of Mathematics}.\hskip 1em plus 0.5em minus 0.4em\relax Institute
  for Advanced Study: \url{https://homotopytypetheory.org/book}, 2013.

\bibitem{Riehl2017}
E.~Riehl and M.~Shulman, ``A type theory for synthetic {$\infty$}-categories,''
  \emph{Higher Structures}, vol.~1, no.~1, pp. 147--224, November 2017.

\bibitem{Abadi1989}
M.~Abadi, L.~Cardelli, P.-L. Curien, and J.-J. Levy, ``Explicit
  substitutions,'' in \emph{Proceedings of the 17th ACM SIGPLAN-SIGACT
  Symposium on Principles of Programming Languages}, ser. POPL '90.\hskip 1em
  plus 0.5em minus 0.4em\relax ACM, 1990, pp. 31--46.

\bibitem{Ritter1997}
E.~Ritter and V.~de~Paiva, ``On explicit substitutions and names (extended
  abstract),'' in \emph{Automata, Languages and Programming}, P.~Degano,
  R.~Gorrieri, and A.~Marchetti-Spaccamela, Eds.\hskip 1em plus 0.5em minus
  0.4em\relax Springer Berlin Heidelberg, 1997, pp. 248--258.

\bibitem{CloneBookRef}
B.~Plotkin, \emph{Universal Algebra, Algebraic Logic, and Databases}.\hskip 1em
  plus 0.5em minus 0.4em\relax Springer, 1994.

\bibitem{Gordon1995}
R.~Gordon, A.~Power, and R.~Street, \emph{Coherence for tricategories}.\hskip
  1em plus 0.5em minus 0.4em\relax Memoirs of the American Mathematical
  Society, 1995.

\bibitem{Gurski2013}
N.~Gurski, \emph{Coherence in Three-Dimensional Category Theory}.\hskip 1em
  plus 0.5em minus 0.4em\relax Cambridge University Press, 2013.

\bibitem{agda}
{Agda contributors}, ``The {A}gda proof assistant,''
  {\url{https://wiki.portal.chalmers.se/agda/pmwiki.php}}.

\bibitem{Power1998}
A.~J. Power, ``2-categories,'' \emph{BRICS Notes Series}, 1998.

\bibitem{cfwm}
S.~Mac~{L}ane, \emph{Categories for the Working Mathematician}, 2nd~ed., ser.
  Graduate Texts in Mathematics.\hskip 1em plus 0.5em minus 0.4em\relax
  Springer-Verlag New York, 1998, vol.~5.

\bibitem{TFiore2006}
T.~Fiore, \emph{Pseudo Limits, Biadjoints, and Pseudo Algebras: Categorical
  Foundations of Conformal Field Theory}, ser. Memoirs of the American
  Mathematical Society.\hskip 1em plus 0.5em minus 0.4em\relax AMS, 2006.

\bibitem{Lawvere1996}
F.~W. Lawvere, ``Adjoints in and among bicategories,'' in \emph{Logic \&
  Algebra}, ser. Lecture Notes in Pure and Applied Algebra, vol. 180, 1996, pp.
  181--189.

\bibitem{Lack2012}
S.~Lack and R.~Street, ``Skew monoidales, skew warpings and quantum
  categories,'' \emph{Theory and Applications of Categories}, 2012.

\bibitem{Street1980}
R.~Street, ``\BIBforeignlanguage{eng}{Fibrations in bicategories},''
  \emph{\BIBforeignlanguage{eng}{Cahiers de Topologie et G{\'e}om{\'e}trie
  Diff{\'e}rentielle Cat{\'e}goriques}}, vol.~21, no.~2, pp. 111--160, 1980.

\bibitem{Staton2013}
S.~Staton, ``An algebraic presentation of predicate logic,'' in
  \emph{Foundations of Software Science and Computation Structures},
  F.~Pfenning, Ed.\hskip 1em plus 0.5em minus 0.4em\relax Springer Berlin
  Heidelberg, 2013, pp. 401--417.

\bibitem{Fiore2017}
M.~Fiore, ``On the concrete representation of discrete enriched abstract
  clones,'' \emph{Tbilisi Mathematical Journal}, vol.~10, no.~3, pp. 297--328,
  2017.

\bibitem{Szlachanyi2012}
K.~Szlach{\'a}nyi, ``Skew-monoidal categories and bialgebroids,''
  \emph{Advances in Mathematics}, vol. 231, no.~3, pp. 1694--1730, 2012.

\bibitem{Leinster2004}
T.~Leinster, \emph{Higher operads, higher categories}, ser. London Mathematical
  Society Lecture Note Series.\hskip 1em plus 0.5em minus 0.4em\relax Cambridge
  University Press, 2004, no. 298.

\bibitem{Hermida2000}
C.~Hermida, ``Representable multicategories,'' \emph{Advances in Mathematics},
  vol. 151, no.~2, pp. 164--225, 2000.

\bibitem{Crole1994}
R.~L. Crole, \emph{Categories for Types}.\hskip 1em plus 0.5em minus
  0.4em\relax Cambridge University Press, 1994.

\bibitem{Carboni1987}
A.~Carboni and R.~Walters, ``Cartesian bicategories {I},'' \emph{Journal of
  Pure and Applied Algebra}, vol.~49, no.~1, pp. 11--32, 1987.

\bibitem{Carboni2008}
A.~Carboni, G.~Kelly, R.~Walters, and R.~Wood, ``Cartesian bicategories {II},''
  \emph{Theory and Applications of Categories}, vol.~19, no.~6, pp. 93--124,
  2008.

\end{thebibliography}

\end{document}